\documentclass[twocolumn]{aastex631}

\usepackage{mathtools}
\usepackage{amsmath,esint}

\shorttitle{Coriolis Effects in Stellar Flares}
\shortauthors{Seligman et al.}

\graphicspath{{./}{figures/}}

\begin{document}

\title{Theoretical and Observational Evidence for Coriolis Effects in Coronal Magnetic Fields Via Direct Current Driven Flaring Events }

\correspondingauthor{Darryl Z. Seligman}
\email{dzseligman@uchicago.edu}

\author[0000-0002-0726-6480]{Darryl Z. Seligman}
\affiliation{Department of the Geophysical Sciences, University of Chicago, Chicago, IL 60637, USA}

\author[0000-0003-0638-3455]{Leslie~A.~Rogers}
\affiliation{Department of Astronomy and Astrophysics, University of Chicago, Chicago, IL 60637, USA}

\author[0000-0002-9464-8101]{Adina~D.~Feinstein}
\altaffiliation{NSF Graduate Research Fellow}
\affiliation{Department of Astronomy and Astrophysics, University of Chicago, Chicago, IL 60637, USA}

\author[0000-0003-3893-854X]{Mark R. Krumholz}
\affiliation{Research School of Astronomy and Astrophysics, Australian National University, Canberra, ACT 2611, Australia}
\affiliation{ARC Centre of Excellence for Astronomy in Three Dimensions (ASTRO-3D), Canberra, ACT 2611, Australia}

\author[0000-0001-9199-7771]{James R. Beattie}
\affiliation{Research School of Astronomy and Astrophysics, Australian National University, Canberra, ACT 2611, Australia}

\author[0000-0002-0706-2306]{Christoph Federrath}
\affiliation{Research School of Astronomy and Astrophysics, Australian National University, Canberra, ACT 2611, Australia}
\affiliation{ARC Centre of Excellence for Astronomy in Three Dimensions (ASTRO-3D), Canberra, ACT 2611, Australia}

\author[0000-0002-8167-1767]{Fred~C.~Adams}
\affiliation{Physics Department, University of Michigan, Ann Arbor, MI 48109}
\affiliation{Astronomy Department, University of Michigan, Ann Arbor, MI 48109}

\author{Marco Fatuzzo}
\affiliation{Physics Department, Xavier University, Cincinnati, OH 45207}

\author[0000-0002-3164-9086]{Maximilian~N.~G{\"u}nther}
\altaffiliation{ESA Research Fellow}
\affiliation{European Space Agency (ESA), European Space Research and Technology Centre (ESTEC), Keplerlaan 1, 2201 AZ Noordwijk, The Netherlands}%

\begin{abstract}
All stars produce explosive surface events such as flares and coronal mass ejections. These events are driven by the release of energy stored in coronal magnetic fields, generated by the stellar dynamo. However, it remains unclear if the energy deposition in the magnetic fields is driven by direct or alternating currents. Recently, we presented observational measurements of the flare intensity distributions for a sample of $\sim10^5$ stars across the main sequence observed by \textit{TESS}, all of which exhibited power-law distributions similar to those observed in the Sun, albeit with varying slopes. Here we investigate the mechanisms required to produce such a distribution of flaring events via direct current energy deposition, in which coronal magnetic fields braid, reconnect, and produce flares. We adopt a topological model for this process which produces a power-law distribution of energetic flaring events. We expand this model to include the Coriolis effect, which we demonstrate produces a shallower distribution of flare energies in stars that rotate more rapidly (corresponding to a weaker decline in occurrence rates toward increasing flare energies). We present tentative evidence for the predicted rotation-power-law index correlation in the observations.  We advocate for future observations of stellar flares that would improve our measurements of the power-law exponents, and yield key insights into the underlying dynamo mechanisms that underpin the self-similar flare intensity distributions.
\end{abstract}

\keywords{emerging flux tubes -- solar dynamo -- solar magnetic reconnection -- optical flares}

\section{Introduction} \label{sec:intro}

Since the first recorded observation of a Solar superflare in 1859 by \citet{Carrington1859}, the mechanisms driving explosive flaring in stars have been the subject of detailed inquiry from plasma physicists and astronomers alike. The origin of flares is closely related to the still-open problem of what mechanism is responsible for heating the Solar corona \citep{Withbroe1977}. In the working picture that has emerged since 1859, it has become clear that magnetic fields are an important component of the heating \citep{Golub1997,Parker1989,Schrijver2000}.
The concentration and twisting of magnetic field lines can lead to the release of energy via the process of magnetic reconnection in flaring events  \citep{Sturrock1984,Kulsrud1998,Priest2000}. In this study, we are interested in understanding the role of the Coriolis force in the dynamo of other stars, and to accomplish this we apply theoretical insights about the heating of the Solar corona and the energy mechanisms driving Solar flares to other stars. 

The role of the Coriolis force in the Solar dynamo  produces observational signatures in the inclination, twist and location of coronal loops and active regions.  Loops tend to be inclined with respect to the latitudinal line \citep{Hale1919,Howard1993,Choudhuri1987,Choudhuri1990,Howard1993,DSilva1993,Longcope1996,Longcope1998,Weber2011,Weber2013}. Magnetic fields in active regions tend to twist in the opposite direction in the North and South hemisphere \citep{Seehafer1990,Pevtsov1995,Abramenko1997,Bao1998,Longcope2000,Longcope2003,Seligman2014,Manek2018,Manek2021}. Faster rotating sunspots are more likely to be towards the equator and slower rotating spots towards the pole \citep{DSilva1995}. Stars spin down as they age \citep{Noyes1984,Soderblom2010,Notsu2013,Candelaresi2014,Doyle2018,ilin19,Doyle2019,Doyle2020}. Recent work has considered the role of the coriolis force in the dynamo of other stars with different rotation rates \citep{Holzwarth2007,Kitchatinov2015,Weber2016,Browning2016,Solanki2006,Parker2009,Rempel2011}. In this paper we consider an additional signature of the Coriolis force: its impact on the distribution of flares.

Observations at different wavelengths have demonstrated that the distribution of peak intensity of Solar flares follow power-laws \citep{Drake1971,Datlowe1974,Dennis1985,Lin1984}, where

\begin{equation}\label{eq:distribution}
  \frac{dN}{dE}\sim E^{-\alpha}\,,
\end{equation}
and $N$ is the number of flares, $E$ is the total energy released in the flare (where the flare magnitude $m\propto\log E$), and the power-law exponent falls roughly in the range of $\alpha\sim1.25-1.5$. \citet{Wheatland2000} found that typical flare frequency distribution (FFD) indices did not vary for individual active regions on the Sun. However, it is important to note that converting from the intensity to energy requires assumptions about the geometry and physical conditions of the flaring region, the mechanisms producing extremely energetic photons \citep{Lee1993,Brown1998}, and the flare height as a function of amplitude \citep{MitraKraev2001}.

A useful parameter to describe the potential mechanisms driving coronal heating and the coupling of the convective interior and the corona  is the plasma $\beta$ parameter,

\begin{equation}
    \beta=2\frac{c_s^2}{v_A^2}\,,
\end{equation}
where $c_s$ is the sound speed and $v_A$ is the Alfv\'en speed. When $\beta>1$ acoustic modes dominate the dynamics and transport of energy, and when $\beta<1$ Alfv\'enic modes dominate. This is useful for identifying the importance of electrodynamic coupling relative to mechanical coupling between the inner and outer atmosphere of the Sun. The electrodynamic coupling will dominate if $\beta<1$ in the outer atmosphere and $\beta\sim1$ in the inner atmosphere, where mechanical dynamics such as convection or differential rotation can couple to Alfv\'enic perturbations in the corona and drive the heating \citep{Ionson1985}. This is generally believed to be the case for the Sun, and similarly for other stars.

Two broad mechanisms that have been proposed to explain the heating of the Solar corona \citep{Withbroe1977} are direct current (DC) and alternating current (AC). Magnetic stress or DC heating dominates when large scale subsurface fluid motions have timescales that are much longer than the Alfv\'enic  coronal crossing timescale, $\tau_A=L_c/v_A$ (where $L_c$ is the radial extent  of the corona) \citep{Ionson1982}, while AC heating dominates in the opposite limit. 

The fundamental idea behind DC models originates in the work of \citet{Parker1972}, who demonstrated that a large scale magnetic field could not be in hydrostatic equilibrium if field lines were topologically braided or knotted.  The nonexistence of an equilibrium is robust to any pressure perturbations applied along the field line. Parker concluded that braided or knotted flux tubes produce rapid dissipation and merging of field lines into a one-dimensional topology. \citet{Parker1983} connected this simple model to heating of the Solar corona via DC currents, whereby slow random walks in the footpoint positions generate  magnetic braids that must reconnect and release energy. This braiding mechanism for heating of the Solar corona was extended by many authors \citep{mikic1989,Berger1993,Parker1994,Longcope1994,Galsgaard1996,Berger2009,Berger2015}.

In the DC regime, a simplified view is that reconnection is triggered when the  angle, $\theta$, between neighboring magnetic field vectors is greater than a critical angle $\theta_c$, which may trigger  smaller events in neighboring field lines \citep{Sturrock1984,Porter1987,Parker1988,Sturrock1990,Berger1993,Krucker2000}, sometimes referred to as nanoflares. These were recently observed by \citet{Antolin2021}, who verified that these could be explained as  reconnection events at small angles. \citet{Parker1988} conjectured that $\theta_c\sim30^\circ$, although \citet{Dahlburg2005} argued that $\theta_c\sim45^\circ$ with a more detailed analysis that included secondary instability. It is important to note, however, that models of coronal heating such as that presented by \citet{vanBallegooijen1986} required reconnection events even at small $\theta_c$ in a cascade of magnetic energy transport. This model was furthered by \citet{Cargill1994}, who presented a model of an active region as hundreds of small elemental flux loops randomly heated by nanoflares. In any case, when $\theta<\theta_c$, magnetic reconnection proceeds slowly, and magnetic energy is deposited in the form of braided fields  quadratically in time \citep{Parker1983,Moffatt1990,Berger1993}. The combination of twisting and braiding of field lines could explain the flare frequency spectrum observed in active regions \citep{Zirker1993i,Zirker1993ii}.

For AC models, the timescale of the footpoint motions is of order or shorter than the coronal crossing timescale, so footpoint motions excite Alfv\'enic modes that travel back and forth along coronal magnetic field lines \citep{vanBallegooijen2011,Asgari2012,Asgari2013,vanBallegooijen2014}. This regime is  driven by smaller scale surface convection mechanisms  \citep{Ionson1985}. \citet{Asgari2014} demonstrated that the AC model was consistent with observations of  non-thermal widths of coronal emission of Fe~\textsc{xii}, Fe~\textsc{xiii}, Fe~\textsc{xv}, and Fe~\textsc{xvi} from the Extreme-ultraviolet Imaging Spectrometer on the \textit{Hinode} spacecraft. However, it is not clear that AC models can explain the existence of exceptionally hot $>5$ MK coronal loops \citep{Asgari2015}, and it has been speculated that  DC  events are important there.

The heating of the Solar corona has been linked to the theory of self-organized criticality \citep{bak1987,bak1988}, which describes dissipative dynamical systems that remain in a critical state with no intrinsic length or time scale. The theory requires a local instability mechanism that can trigger neighboring instabilities.  This avalanching mechanism produces energetic events at all length scales \citep{Kadanoff1989,Babcock1990}. Applications of the theory have been hypothesized in turbulence, percolation systems \citep{Turcotte1999},  neuroscience \citep{Ribeiro2010,Hesse2014}, landslides \citep{Bak1990,Turcotte2002}, atmospheric dynamics \citep{Grieger1992,Andrade1998},  astrophysical accretion disks \citep{Dendy1998}, traffic patterns \citep{Nagel1993}, evolution \citep{Bak1993}, extinction events \citep{Newman1996_criticality}, financial markets \citep{Bak1997}, earthquakes \citep{Gutenberg1956,bak1989_earthquakes,Sornette1989,Olami1992,Carlson1989}, and Conway's game of Life \citep{bak1989}. \citet{lu1991} proposed that the Solar coronal magnetic field is also in a self-organized critical state to explain the power-law observed in the magnitude of Solar flares \citep{Lu1993,Crosby1993,Aschwanden1998,Charbonneau2001,deArcangelis06}.  This approach has been powerful in prediction of extreme flares \citep{Morales2020}.

In the DC picture of coronal heating, the slow buildup of braided fields  provides all of the requisites for a self-organized critical system. Twisted coronal fields are generated via dynamo mechanisms in the fluid-dominated interior \citep{Charbonneau2010},  via convective and Coriolis driven vortical subsurface plasma flows \citep{Parker1955,Moffat1978,Longcope1998,Seligman2014}.   \citet{Prior2016} suggested that active Solar regions could be formed via the injection of rising magnetic field topologies that were only braided, and not twisted.  \citet{MacTaggart2021} demonstrated numerically that active regions are created by the emergence of a large flux tube of pre-twisted magnetic fields by examining the evolution of the topological quantity magnetic winding in the emergence of active regions.  \citet{Berger2009} and \citet{Berger2015} demonstrated that  a topological model that included braiding and reconnection of coronal fields  exhibited power-law distributions of energetic events.

It is important to note that while there exist physical connections between DC heating models and  self-organized criticality, the theory is not the only way to create a power-law distribution of energetic events \citep{Rosner1978,Litvinenko1996}. For example, \citet{Newman1996} demonstrated a dynamical system driven by coherent noise could arrive at a similar stationary state characterized by power-law distributions of avalanches, but without maintaining the ``critical" state. MHD turbulence invoked in the AC regime  can produce flare-like  energy occurrence distributions without relying on the theory of self-organized criticality \citep{Longcope1994,Einaudi1996,Galsgaard1996,Dmitruk1997,Galtier1998,Georgoulis1998,Einaudi1999,Galtier1999}.

To date, most studies of heating and flaring mechanisms have focused on the Sun. While this is obviously the system for which we can obtain the richest and highest-quality data, it remains a sample of one. Studies of flaring in other stars can therefore offer a unique and valuable perspective. Flare-like X-ray emissions from other stars exhibit power-law energy distributions with similar indices to those seen in the Sun \citep{Shakhovskaia1989,Osten1999,Audard2000}. \citet{Aschwanden21} found a power-law dependence of energies for optical flares observed  with \textit{Kepler} \citep{Davenport16}. The \textit{TESS} mission provided 2-minute cadence light curves for $\sim$200,000 stars, which allowed for the identification of a statistically significant sample of flaring events \citep{guenther19_flares}. \citet{Feinstein2021} demonstrated that the these flares follow a power-law distribution of intensity (with slopes $\alpha' \approx 0.9 - 1.5$) for all main sequence stars.\footnote{$\alpha'$ indicates the slope analogous to $\alpha$ in Equation \ref{eq:distribution}, for the normalized intensity distribution.} These newly measured values are close to the median of previously measured slopes \citep[Figure 3 in \citet{Feinstein2021}; ][]{shibayama13, guenther19_flares, ilin19, lin2019, howard_ward19, yang19, feinstein20, raetz20, ilin21, Aschwanden21}. In this paper, we attempt to interpret these power-law distributions under the framework of the DC heating mechanism. 

This paper is organized as follows. In \S\ref{sec:braiding_review}, we review the DC  braiding model of reconnection events presented by \citet{Berger2009} and \citet{Berger2015}. We expand the braiding model to include a Coriolis-driven bias in handedness of braids injected, and calculate the resulting distributions of energetic events. In \S\ref{sec:observations}, we present observations of flare frequency distributions (FFDs) for slow and fast rotating stars, and show that they are consistent with the analytic predictions. In \S\ref{sec:conclusions} we conclude and outline future observational and theoretical work.

\section{Generalised Topological Braiding Model}\label{sec:braiding_review}

Reconnection is at its base a phenomenon whereby (anomalous) resistivity dissipates the currents that sustain magnetic fields in a plasma, leading to a violation of flux freezing and allowing rearrangement of the magnetic topology. Regions where reconnection occurs must have a current flowing through them (since otherwise there would be nothing to dissipate), and thus necessarily have
\begin{equation}
    |\,\vec{\nabla}\times \vec{B}\,| \neq 0\,.
\end{equation}
This result follows from consideration of Amp\'ere's Law,
\begin{equation}
    \vec{\nabla}\times\vec{B} = \frac{1}{c}\left(4\pi\vec{J} + \frac{\partial\vec{E}}{\partial t}\right)\,.
\end{equation}
In a plasma that is overall electrically neutral, the electric field in the rest frame vanishes, $|\vec{E}| = 0$, so that $\partial\vec{E}/\partial t = 0$. It is worth noting that in non-inertial frames,  $\vec{E} = \vec{B}\times \vec{V}/c  $, where $\vec{V}$ is the plasma velocity and c is the speed of light. However, $\partial\vec{E}/\partial t <<1$, because in  a non-relativistic plasma, $|\vec{V}|/c<<1  $. The presence of a current, $|\vec{J}| \neq 0$, thus implies a non-vanishing curl in the magnetic field, $|\vec{\nabla}\times\vec{B}| \neq 0$. This condition, in turn, implies that the reconnection rate --- and the flaring rate --- must be connected to the distribution of the curl of the magnetic field. This insight motivates the analysis that follows. 

In this section, we review the braiding model for DC coronal heating via reconnection events presented in \citet{Berger2009} and further developed in \citet{Berger2015}. We then generalise the braiding model to include a Coriolis-induced handedness bias. Before we begin, it is worth noting that other authors have investigated the role of helicity and braiding in active regions and found that both are important for supplying energy to the corona \citep{Longcope2007,Liu_2014_helicity_preference}. 

\subsection{Estimate of Free Magnetic Energy}

We consider a small patch of a stellar atmosphere. We define height along the $\hat z$ direction, defined such that $z=0$ is the base of the fields and $z=L$ represents some radial extent into the corona. At $z=0$, which is nominally at the photosphere, although the exact position is irrelevant, subsurface convection provides stochastic forcing of the position for the magnetic flux tubes.   

As was introduced in Section~2 of \citet{Berger2009}, we consider a braid consisting of $N$ flux tubes or magnetic field lines that permeate the volume. The set of flux tubes is characterized by the number of crossing points, $C$, at which magnetic flux tubes cross. Each flux tube has a number of crossings, $n_{\rm C}$, on average, given by

\begin{equation}
    n_{\rm C} = \bigg( \frac{2 C}{N}\, \bigg)\,,
\end{equation}
so the typical height between crossing junctures is

\begin{equation}
    \delta z = \frac{L}{n_{\rm C}}\,,
\end{equation}
as was demonstrated in Equation~(1) of \citet{Berger2009}. Let $D$ be the typical diameter of a flux tube. Note that this quantity is related to typical separations between magnetic field lines, and is given by the resistivity of the ambient medium. If two of the tubes wrap around each other between heights of $z$ and $z+\delta z$, the distance that they travel in the horizontal direction is roughly $\delta\ell=\pi D/2$. Therefore the ratio between the perpendicular and parallel components  of the magnetic field with respect to the radial direction, $B_\bot$ and $B_\parallel$, is approximately given by

\begin{equation}
   \bigg(\, \frac{\delta \ell}{\delta z}\bigg)\,\sim\bigg(\, \frac{B_\bot}{B_\parallel}\bigg)\,\sim \bigg(\,\frac{\pi D}{NL}\,\bigg)\,C\,.
\end{equation}
This step is important, because it demonstrates that the perpendicular magnetic field is directly proportional to the number of crossings in a flux tube. \citet{Berger1993} demonstrated that the free magnetic energy density is proportional to the square of the perpendicular or transverse magnetic field, so that the free energy per unit volume of the braided field obeys the scaling 
\begin{equation}\label{eq:efree}
    E_{\rm Free}\propto C^2\,,
\end{equation}
as given by Equation~(5) of \citet{Berger2009}.

\begin{figure*}[]
\begin{center}
       \includegraphics[scale=0.5,angle=0]{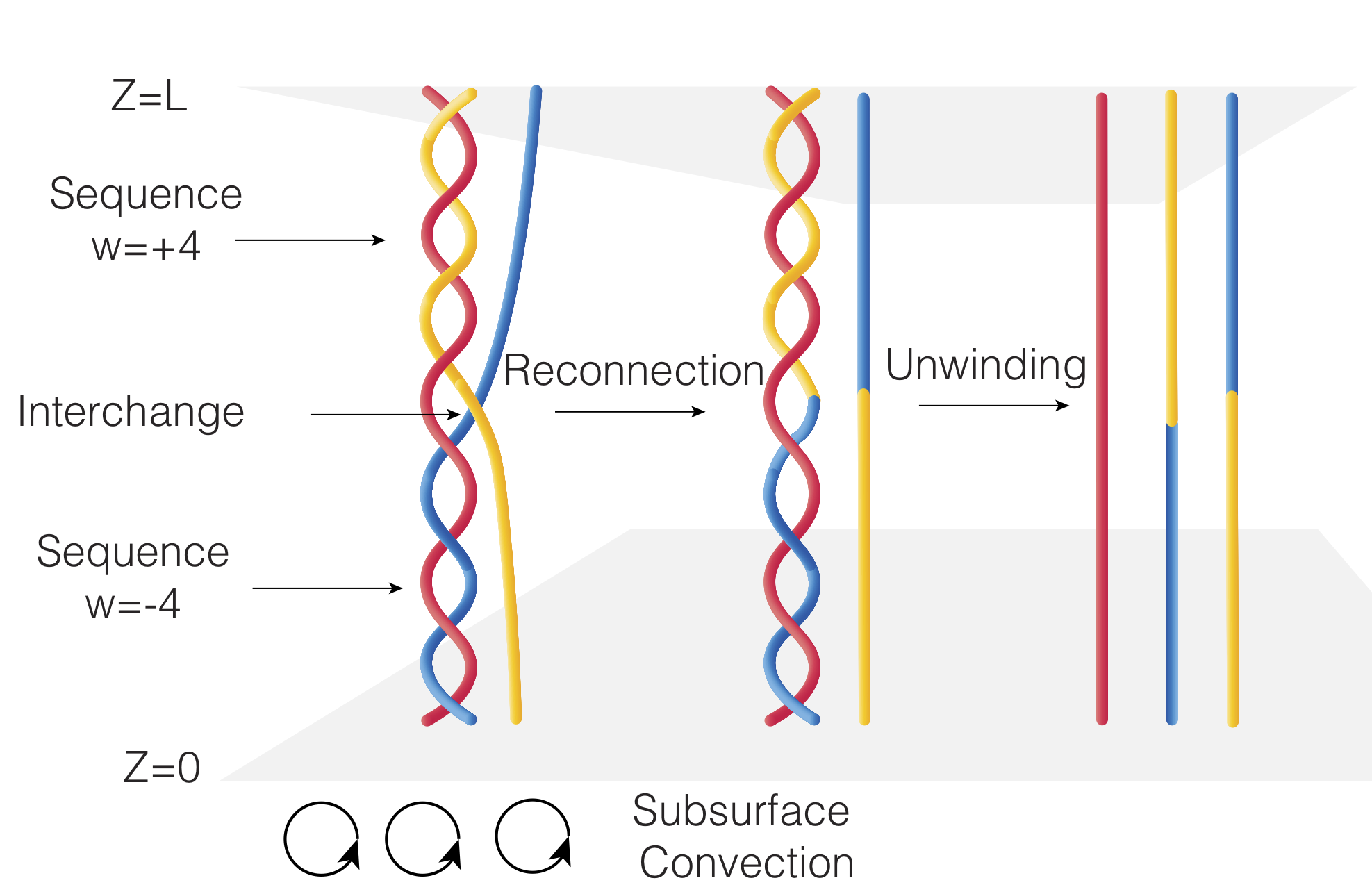}
    \caption{A schematic diagram of three braided flux tubes that cause an incoherent  reconnection  event - resulting in the merging of opposite signed sequences - that releases energy. In this setup, the two sequences on the left hand side are braided in opposing directions. After the reconnection event, which removes the interchange, the two sequences unwind and release the magnetic energy density that was stored in transverse components of the fields. This figure is similar to Figure~4 in \citet{Berger2009}.}\label{fig:schematic_braiding} 
\end{center}
\end{figure*}

\subsection{Evolution to Equilibrium}

\citet{Berger2009} considered a topological model of the braiding of flux tubes to explain the distribution of energy in reconnection events observed in the Solar corona. In the model, there are two defining topological quantities for a nest of individual magnetic flux tubes. The first is the \textit{winding} number $w$ of an individual sequence describing the braiding of two flux tubes around each other. This quantity encodes both the number of crossings, where $|w|=C$, and the handedness of the braiding itself. An \textit{interchange} represents a region where a third flux tube crosses one of the two original flux tubes, and is inserted into the original braiding strand. These interchanges are the sites of reconnection events. 

In Figure~\ref{fig:schematic_braiding}, we present a schematic diagram of the topological model, similar  to Figure~4 in \citet{Berger2009}. The left hand side shows a braid consisting of two sequences and one interchange. The primary flux tube is shown in red, and takes part in both of the sequences. The blue flux tube is braided around the red flux tube from the base where $z=0$, and these two create a sequence with $w_1=-4$, of negative handedness. Above the sequence, the blue flux tube is removed from the braid in an interchange with the yellow flux tube. The yellow flux tube then creates a sequence where $w_2=+4$ with the red flux tube above the interchange, of positive handedness. 

\citet{Berger2009} considered a model in which the coronal magnetic field consisted of many of these types of braided flux tubes, nested together. The defining feature of the model is that a reconnection event is modeled as the removal of an interchange. This process is demonstrated in the middle portion of Figure~\ref{fig:schematic_braiding}, where the interchange is removed and the blue and yellow flux tubes merge on the right hand side. After the removal of the interchange, the two sequences above and below the interchange are able to merge. The final winding, $w_f$ of the resulting single sequences is given by $w_f=w_1+w_2$. In the configuration shown in the schematic figure, the equal and opposite windings above and below the interchange cancel each other out completely. In other words, the entire braid completely unwinds itself. This behavior is shown on the right hand side of the figure, where the three flux tubes are now relaxed to an equilibrium where there is no braiding. Because of the results in the previous subsection, and namely Equation~(\ref{eq:efree}), the free magnetic energy that is released in the reconnection event encoded by the removal of the interchange is given by the number of winds that are undone. 

Remarkably, \citet{Berger2009} presented a method to evolve this simplistic model through time to an equilibrium state, which we summarize here. At some arbitrary initial time $t=0$, consider a neighborhood in the corona containing an integer number of braiding sequences, $\Lambda$, which by construction contains  $\Lambda-1 $  interchanges. The system then evolves through a series of idealised discrete time steps. During any given time step,
\begin{itemize}
    \item 1 sequence is added,
    \item  1 interchange is added,
    \item  1 reconnection event occurs, removing 1 pre-existing interchange. Sequences on either side of the interchange merge.
\end{itemize}
\citeauthor{Berger2009} then investigate what steady-state distribution of windings $w$ such an evolutionary sequence produces. At any point, the number of sequences where $C=w$ is given by $\xi(w)$. Therefore,
\begin{equation}
    \Lambda = \sum_{w=-\infty}^\infty \xi(w)\,,
\end{equation}
as defined with different notation in Equation~(7) of \citet{Berger2009}. The probability that there will be a sequence of length $w$ is given by the normalized probability function, $f(w)$, which is defined in Section 5.2 of \citet{Berger2009} as 
\begin{equation}
    f(w)=\frac{\xi(w)}{\Lambda}\,.
\end{equation}

Next, the probability of adding a new sequence of $w$ at each time step is given by a function defined to be $p(w)$. At each time step in the algorithm, the probability function evolves to a new value through the mapping 

\begin{equation}
    f(w)\rightarrow f(w) +\delta f(w)\,,
\end{equation}
 defined by the three events that are allowed to happen in each time step itemized above. Therefore, this process can be described at each time step by the following equation,
\begin{equation}\label{eq:model}
\begin{split}
    \Lambda \delta f(w) & = \delta \xi(w) = p(w) -2 f(w)+\\ 
    & \int_{-\infty}^{\infty}f(\omega_1)d\omega_1\int_{-\infty}^{\infty}f(\omega_2)\delta\big(w-(\omega_1+\omega_2)\big)d\omega_2\,.
\end{split}
\end{equation}
The three terms on the right hand side denote (in order from left to right) the addition of a new sequence, the removal of the surrounding sequences, and the addition of the resulting merged sequence.  
In order to evolve this model to an equilibrium, it is sufficient to set Equation~(\ref{eq:model}) equal to zero to indicate that a  steady state has been reached. When the equilibrium is reached, the mapping no longer changes the probability density function and $\delta \xi(w) = 0$.  At this point, the equation reduces to
\begin{equation}\label{eq:f_p}
    p(w)=2 f(w) - (f \star f)(w)\,,
\end{equation}
where the $\star$ indicates a convolution, as in Equation~(11) of \citet{Berger2009}. It is straightforward to demonstrate, as in Section 5.2 of \citet{Berger2009}, that this can be solved in Fourier space via Fourier transforms and inverse Fourier transforms. Assuming that the input of interchanges is a Poisson process such that the windings inserted in the sequences between interchanges follow an exponential distribution,
\begin{equation}
    p(w)=\frac{\lambda}{2}e^{-\lambda |w|}\,,
\end{equation}
for some constant $\lambda$, then the equilibrium solution exhibits a probability distribution of windings characterized by the equation 
\begin{equation}
    f(w)=\frac{\lambda}{2}\bigg[L_{-1}(\lambda w)-I_1(w\lambda)\bigg]\,,
\end{equation}
where $I_1$ is the Bessel-$I$ function and $L_{-1}$ is a Struve-$L$ function. We note that this function is slightly different than Equation~(17) in \citet{Berger2009}, although we have verified that our solution has the same form as Figure~5 in their paper. Importantly, for winding numbers with magnitude greater than unity, $|w|>1$, the probability distribution has the form of a power-law where $ f(w) \sim w^{-2}$.

\subsection{Energy Distribution of Flares}\label{subsection:energy}

Consider the distribution of reconnection events in the previous subsection. Since we know that the magnetic free energy scales with the square of the crossing number of the set of flux tubes, this must also correspond to a distribution or power law of energetic flaring events once the  equilibrium state has been reached, since $E_{\rm Free}\sim C^2$ as in Equation~(\ref{eq:efree}). Therefore, the difference in energy before and after a reconnection event, $\Delta E$, is given by the difference in the square of the initial and final values of $C$, i.e., 
\begin{equation}
    \Delta E \sim C_0^2 - C_F^2\,.
\end{equation}

\citet{Berger2009} assumed that the reconnection occurs after the winding is greater than some critical value, $C_{\rm crit}$. They assumed the interchange removal merged two sequences with twist numbers $\omega_1$ and $\omega_2$, where by convention $|\omega_2|>|\omega_1|$. Critically, they assume that the  merged sequences have opposite sign, on the basis that mergers of sequences with the same winding direction will not create a release of energy in a flaring event. The merging after interchange produces a single sequence with $\omega_f=|\omega_2|-|\omega_1|$, and energy
\begin{equation}
    \Delta E \sim C_{\rm crit}^2 - \bigg( C_{\rm crit}-2|\omega_1|\bigg)^2\,.
\end{equation}
If $\omega_1\ll C_{\rm crit}
$, then to first order we have
\begin{equation}\label{eq:ecrit}
    E\sim 4 C_{\rm crit} |\omega_1|\,,
\end{equation}
and we can deduce the distribution of flare energies, $F(E)$ from the relationship between energies before and after reconnection and the distribution of winding numbers $w$:
\begin{equation}\label{eq:f_e}
    F(E)\sim E^{-\alpha}\,,
\end{equation}
where $\alpha=2\gamma-1$, and $\gamma$ is the exponent of the winding distribution $f(w)\sim|w|^{-\gamma}$, as demonstrated in Section 5.3 of \citet{Berger2009}.

\subsection{Expansion of Braiding Model}

\begin{figure}[]
\begin{center}
\includegraphics[scale=0.46,angle=0]{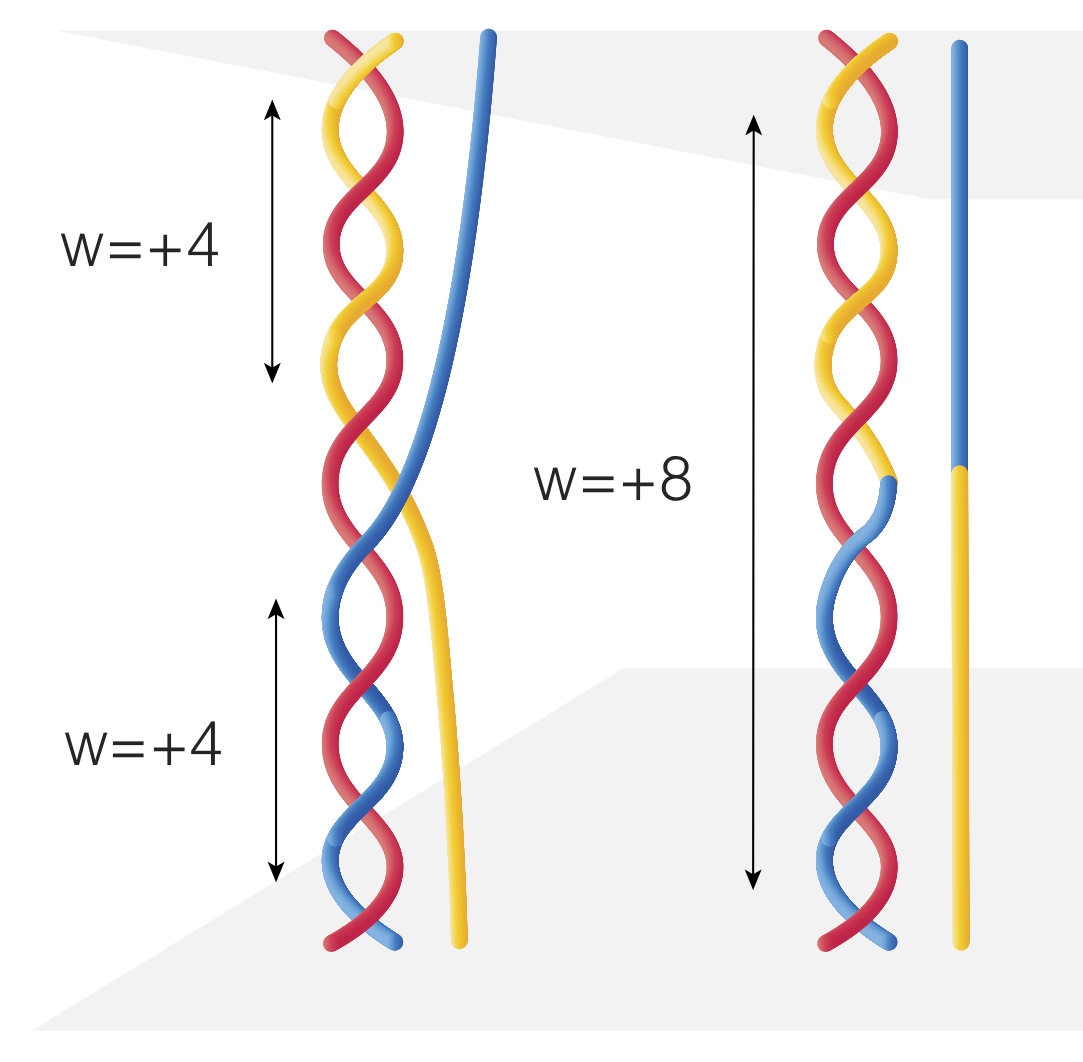}
    \caption{A schematic diagram of a coherent reconnection event. In this setup, the two sequences on the left hand side both have positive winding. After the reconnection event that removes the interchange, the resulting braid on the left hand side has a winding of $w=+8$.}\label{fig:schematic_braiding_constructive} 
\end{center}
\end{figure}

The \citet{Berger2009} model assumes that injection of braids with positive and negative winding are equally probable, which is a reasonable approach for a slowly-rotating star like the Sun, where the Rossby number in the sub-surface convection zone is relatively large \citep{Greer2016}. However, this assumption must begin to fail in more rapidly rotating stars with smaller Rossby numbers, and we must therefore generalize the model to include a bias in the sign of $w$.

To generalize this (but keeping the same distribution on the overall amplitude of $w$ in the injected sequences) we just adjust the prefactor. We define a parameter $\eta\in[0,1]$ to be the probability that a sequence has a positive handed winding number. The probability distribution function for the input of sequences can be written as

 \begin{equation}\label{eq:p_eta_bias}
    p(w) =\begin{dcases}\eta\lambda e^{-\lambda w } & {\rm if} \,\, w>0 \\
    \left(1-\eta\right)\lambda e^{+\lambda w }  \, & {\rm if} \, \,  w<0 \,.
    \end{dcases} 
\end{equation}
In Equation~(\ref{eq:p_eta_bias}), the value of $\eta$ determines the percentage of sequences that are \textit{locally} injected into the corona with the same sign or handedness in the twist. If there is a bias towards more sequences with the same sign, then one would expect that there should be a bias towards more coherent interchange removals, where the surrounding sequences add constructively. 

In Figure~\ref{fig:schematic_braiding_constructive}, we show a schematic in which the two sequences have the same sign and number of windings. After the interchange is removed, only one braid remains, but it has twice the number of windings as the original two braids, so no energy is released from the unwinding of braids. It is natural to expect that as $\eta$ increases or decreases away from $1/2$ (which represents an equal injection of positive and negative sequences), the steady state winding distribution will be skewed towards higher numbers. We can demonstrate this trend analytically by taking the Fourier transform of Equation~(\ref{eq:p_eta_bias}), which takes the form 
 \begin{equation}\label{eq:p_k_eta}
    \tilde{p}(k) =
     \left(\frac{\lambda^2-i\lambda k(2\eta-1)}{\lambda^2+k^2} \right)\, .
\end{equation}
Using this result in Equation~(\ref{eq:f_p}), the solution for the steady state winding distribution function in Fourier space is \begin{equation}\label{eq:f_solution_fourier}
    \tilde{f}(k) =1-\sqrt{\frac{k^2+i\lambda k(2\eta-1)}{\lambda^2+k^2} }\,,
\end{equation}
and the corresponding real-space distribution is
\begin{equation}\label{eq:inverseFFT_eta}
\begin{split}
     f(w)=\frac{1}{2\pi}\int_{-\infty}^\infty\left[1-\sqrt{
    \frac{k^2+i\lambda k(2\eta-1)}{\lambda^2+k^2} }\right] e^{ i k w}dk \,.
\end{split}
\end{equation}
We verified analytically that Equation (\ref{eq:inverseFFT_eta}) is symmetric for positive and negative $w$, such that $f_\eta(w)=f_{1-\eta}(-w)$, and that it is independant of rescaling of $\lambda$, such that for $x=\lambda w$, $f(x)=f(w)/\lambda$. In order to compute this numerically, we decompose Equation (\ref{eq:f_solution_fourier}) into  real and imaginary components,
 \begin{equation}\label{eq:real_ftilde}
\mathfrak{Re}[\tilde{f}(k)]=1-\sqrt{\frac{\left(\phi+k^2\right)}{2\left(k^2+\lambda^2\right)}}
\end{equation}
and
 \begin{equation}\label{eq:imaginary_ftilde}
 \begin{split}
\mathfrak{Im}[\tilde{f}(k)]=-\frac{\lambda k(2\eta-1)}{\sqrt{2\left(k^2+\lambda^2\right)\left(\phi+k^2\right)}}\,,
\end{split}
\end{equation}
where 
\begin{equation}\label{eq:r}
    \phi = \bigg(k^4+k^2\lambda^2(2\eta-1)^2\bigg)^{1/2}.
\end{equation}
It is important to note that $\phi$ in Equation (\ref{eq:r}) is an even function of $k$. Therefore $\mathfrak{Re}[\tilde{f}(k)]$ (Equation [\ref{eq:real_ftilde}]) is an even function of $k$ and $\mathfrak{Im}[\tilde{f}(k)]$ (Equation [\ref{eq:imaginary_ftilde}]) is an odd function of $k$. Equation (\ref{eq:inverseFFT_eta}) then becomes,

\begin{equation}\label{eq:inverseFFT_eta_decomposed}
\begin{split}
     f(w)=\frac{1}{2\pi}\int_{-\infty}^\infty\,\bigg(\,\mathfrak{Re}[\tilde{f}(k)]\,\cos(kw)\,\bigg)\,dk\\-\frac{1}{2\pi}\int_{-\infty}^\infty\,\bigg(\,\mathfrak{Im}[\tilde{f}(k)]\,\sin(kw)\,\bigg)\, dk \,,
\end{split}
\end{equation}
since the imaginary component is an odd function of $k$ and integrates to 0. 

\begin{figure}[]
\begin{center}
      \includegraphics[scale=0.4,angle=0]{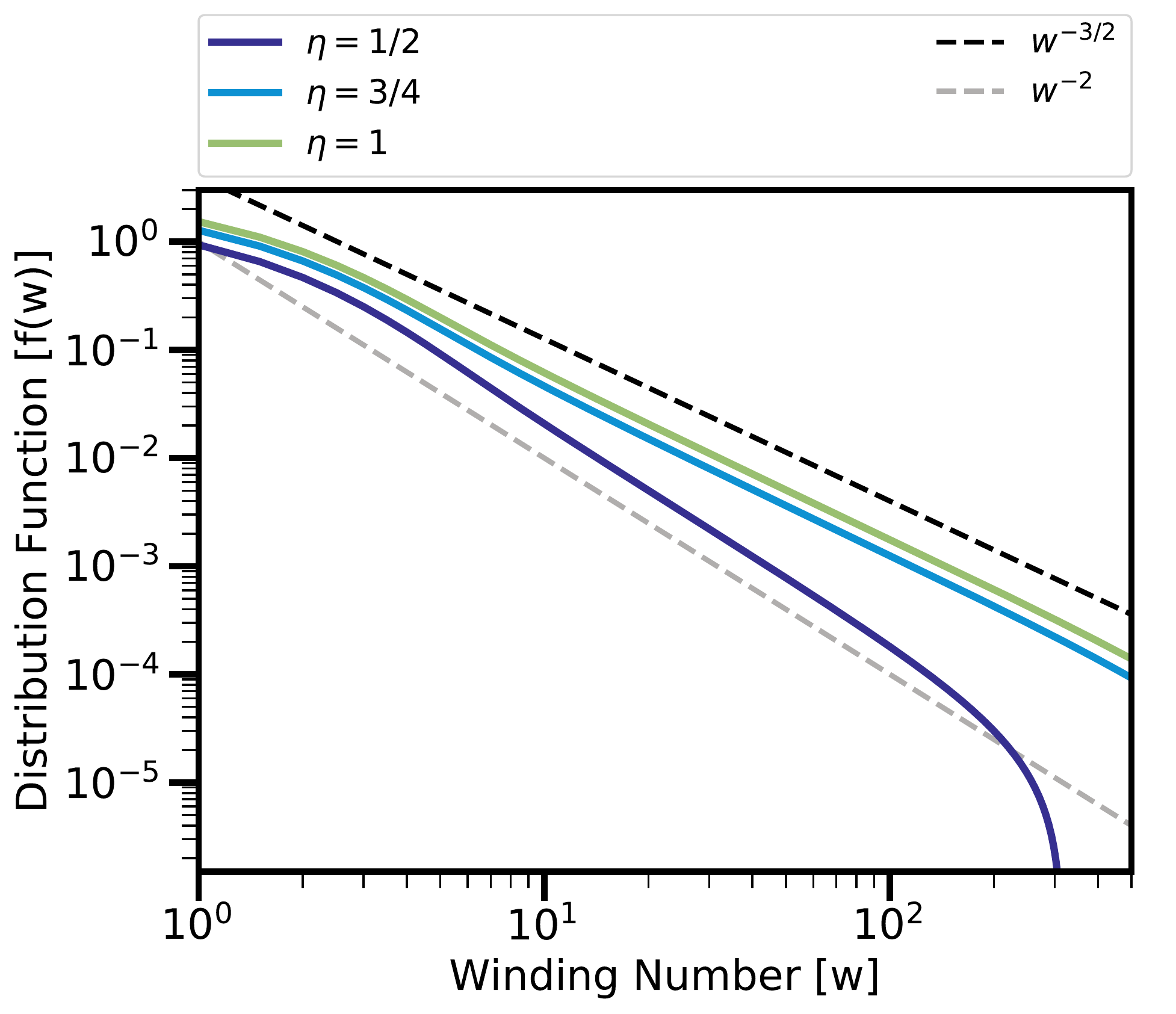}
      \includegraphics[scale=0.4,angle=0]{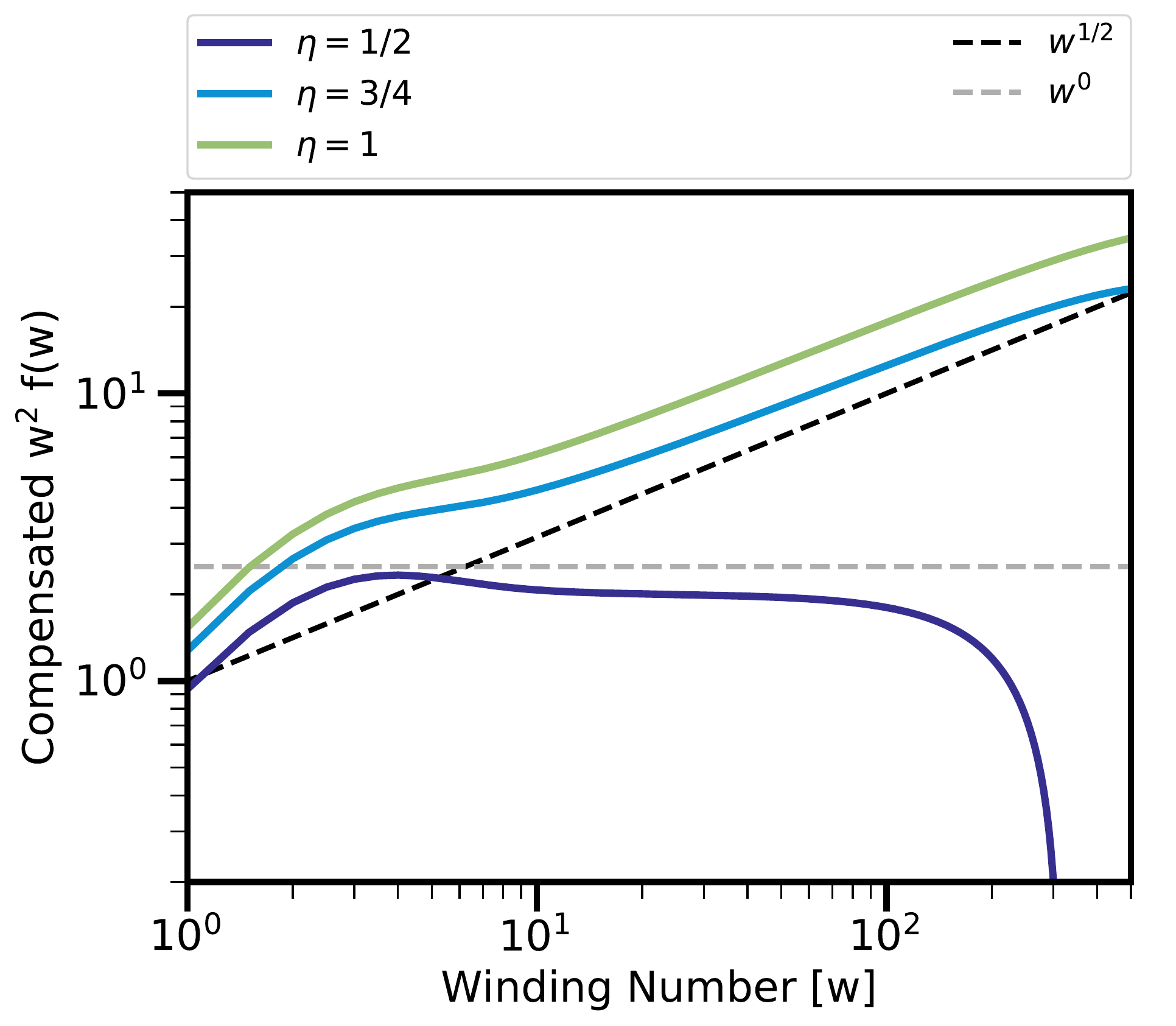}
    \caption{The distribution function for winding of sequences in an equilibrium state. We calculate these distributions by numerically performing the inverse Fourier transform defined by Equation~\ref{eq:inverseFFT_eta} for $\eta=1/2,3/4$ and 1. The purple line shows the solution when $\eta=1/2$ for an equal probability of positive and negative twists, which has a power-law slope of $\gamma=2$, consistent with the findings of \citet{Berger2009}. As $\eta$ increases, and more coherent and same signed braids are injected into the corona, the slopes of the distribution functions get shallower and approach $\gamma=3/2$. This is plotted using $\lambda=1$, where $|w|=1/\lambda$ is the mean  winding magnitude per injected sequence. Changing $\lambda$ simply re-scales $w$, leaving the power-law dependencies unchanged. }\label{fig:powerlaw_eta}  
\end{center}
\end{figure}

\begin{figure}[]
\begin{center}
      \includegraphics[scale=0.4,angle=0]{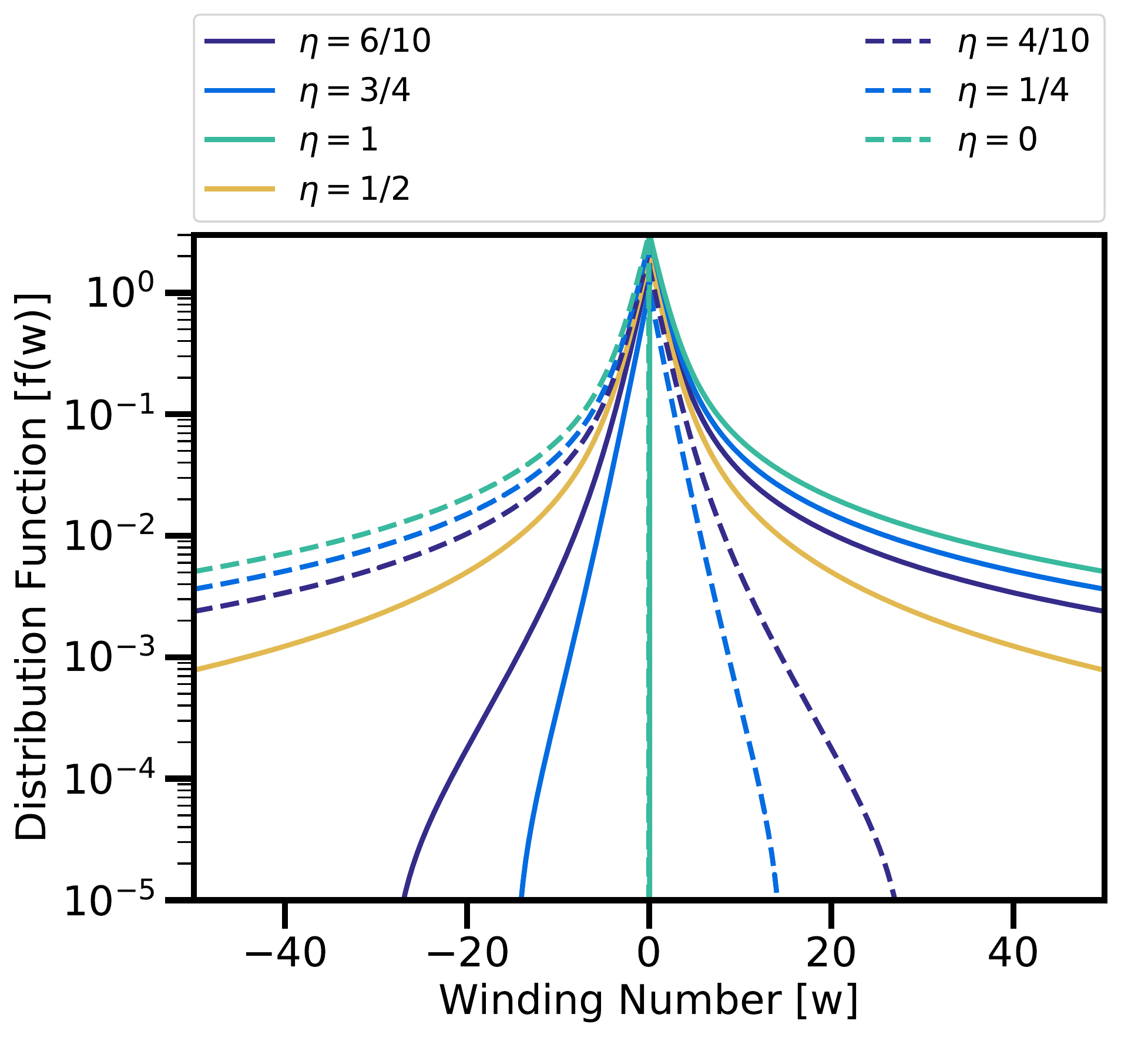}
    \caption{The distribution function for winding of sequences in equilibrium. This is calculated in the same way as in Figure \ref{fig:powerlaw_eta}, and  we show the distribution for positive and negative winding numbers with $\eta=0,1/4,4/10,1/2,6/10,3/4$ and 1. The distributions for $\eta$ and $1-\eta$ are symmetric about $w=0$. The  equilibrium state reflects the same sign bias as the input of sequences in Equation \ref{eq:p_eta_bias}. }\label{fig:positive_negative}  
\end{center}
\end{figure}
We evaluate Equation~(\ref{eq:inverseFFT_eta_decomposed})  numerically and show the resulting distributions in Figure~\ref{fig:powerlaw_eta} for $\eta=1/2,3/4$ and 1. For the case of $\eta=1/2$, the problem reduces to the analytic solution presented in the previous section and in \citet{Berger2009}. As $\eta$ approaches 1 (or 0), and more coherent braids are injected into the corona, the FFD becomes shallower. This result is consistent with the findings of \citet{Berger2009}, who performed Monte Carlo realizations of the problem with an asymmetric injection of positive to negative handed sequences.

The distributions of $f(w)$ in Figure~\ref{fig:powerlaw_eta} are symmetric about the $y$-axis for positive and negative winding numbers when $\eta=1/2$. In Figure~\ref{fig:positive_negative}, we show the equilibrium distributions for positive and negative winding numbers for a range of $\eta\in[0,1]$. When $\eta\ne1/2$, the resulting equilibrium distribution is asymmetric about $w=0$ and contains more sequences with the same sign that is preferentially input. As expected, the equilibrium distributions for $\eta$ and $1-\eta$ (shown with solid and dashed lines) are symmetric about $w=0$ $\forall\eta\in[0,1]$. We note that the end-member case, where $\eta=1$ or $0$, is formally not a physically plausible scenario. This is because, as shown in Figure~\ref{fig:positive_negative}, the distribution functions  for these cases only have sequences with a single sign, which means the star would not be able to produce any energetic flares.

\subsection{Generalized Energy Distribution}
It is important to note that the transformation from the distribution of winding to energy given by Equation \ref{eq:f_e} is no longer valid for asymmetric distributions where $\eta\ne1/2$. Following the notation in Subsection \ref{subsection:energy}, assuming that $|w_2|>|w_1|$ and allowing for two cases where $w_2<0$, $w_1>0$ and $w_2>0$, $w_1<0$, the energy distribution, denoted by capital $F(E)$, presented in Equation 20 in \citet{Berger2009} can be generalized, 

\begin{equation}\label{eq:first_energy}
\begin{split}
F(E)\sim\qquad\qquad\qquad\qquad\qquad\qquad\qquad\qquad\qquad\qquad\\\int_0^\infty\int_{-\infty}^{-w_1}\,\bigg(\,f(w_1)f(w_2)\,\delta\big(E-4|w_1|C_{\rm crit}\big)\,\bigg)\,dw_2dw_1\\+\int_{-\infty}^0\int_{|w_1|}^{\infty}\,\bigg(\,f(w_1)f(w_2)\,\delta\big(E-4|w_1|C_{\rm crit}\big)\,\bigg)\,dw_2dw_1\,.\end{split}
\end{equation}
In this equation, the first term on the right hand side corresponds to the reconnection events where $w_1>0$. Similarly, the second term on the right corresponds to reconnection events where $w_1<0$. Note that the $\delta$ function in both integrands has an absolute value sign because of the form of Equation \ref{eq:ecrit}, where $E\sim|w_1|$. From Figures \ref{fig:powerlaw_eta} and \ref{fig:positive_negative},  the  distribution for $w$ can be approximated as a power law times a negative exponential. The approximate distribution function is given by

 \begin{equation}\label{eq:f_posneg}
    f(w) \sim\begin{dcases}e^{ -(2\eta-1)w/w_+ }|w|^{-\beta} & {\rm if} \,\, w>0 \\
    e^{ (2\eta-1)w/w_- }|w|^{-\psi}  \, & {\rm if} \, \,  w<0 \,,
    \end{dcases} 
\end{equation}
for some truncation winding numbers and power law exponents for the positive and negative cases, $w_+$ and $w_-$, and power law exponents $\beta$ and $\psi$. All four of these parameters depend on the value of $\eta$, and generally $w_+>>w_-$ and $\beta>\psi$ for $\eta\in[1/2,1]$. Equation \ref{eq:f_posneg} was constructed to be symmetric for $\eta$ and $1-\eta$. For the remainder of this subsection, we solve for the energy distribution for the case of $\eta\in[1/2,1]$, since the resulting distribution is the same for $\eta$ and $1-\eta$. By substituting Equation \ref{eq:f_posneg} into Equation \ref{eq:first_energy}, the distribution is given by,

\begin{equation}\label{eq:ugly_integral1}
\begin{split}
F(E)\sim\int_0^\infty\int_{-\infty}^{-w_1}\,\bigg(\,\delta\big(E-4|w_1|C_{\rm crit}\big)\\\,f(w_1)(e^{ (2\eta-1)w_2/w_- }|w_2|^{-\psi})\,\bigg)\,dw_2dw_1+\\\int_{-\infty}^0\int_{|w_1|}^{\infty}\,\bigg(\,\delta\big(E-4|w_1|C_{\rm crit}\big)\,\\f(w_1)(e^{ -(2\eta-1)w_2/w_+ }|w_2|^{-\beta} )\,\bigg)\,dw_2dw_1\,.\end{split}
\end{equation}
This can be solved analytically, by recalling the definition of the incomplete gamma function,

\begin{equation}\label{eq:gamma}
    \Gamma (a,x)= \int_x^\infty t^{a-1}e^{-t}dt\,.
\end{equation}
We perform two change of variables to integrate Equation (\ref{eq:ugly_integral1}). For the first term on the right hand side, we define the variable $v=-(2\eta-1)w_2/w_-$, such that $dw_2=-w_-/(2\eta-1)dv$. Similarly, for the second term on the right hand side, we define the variable $u=(2\eta-1)w_2/w_+$, such that $dw_2=w_+/(2\eta-1)du$. After evaluating the first integral of the double integral in each term, Equation (\ref{eq:ugly_integral1}) reduces to

\begin{figure}[]
\begin{center}
    \includegraphics[scale=0.4,angle=0]{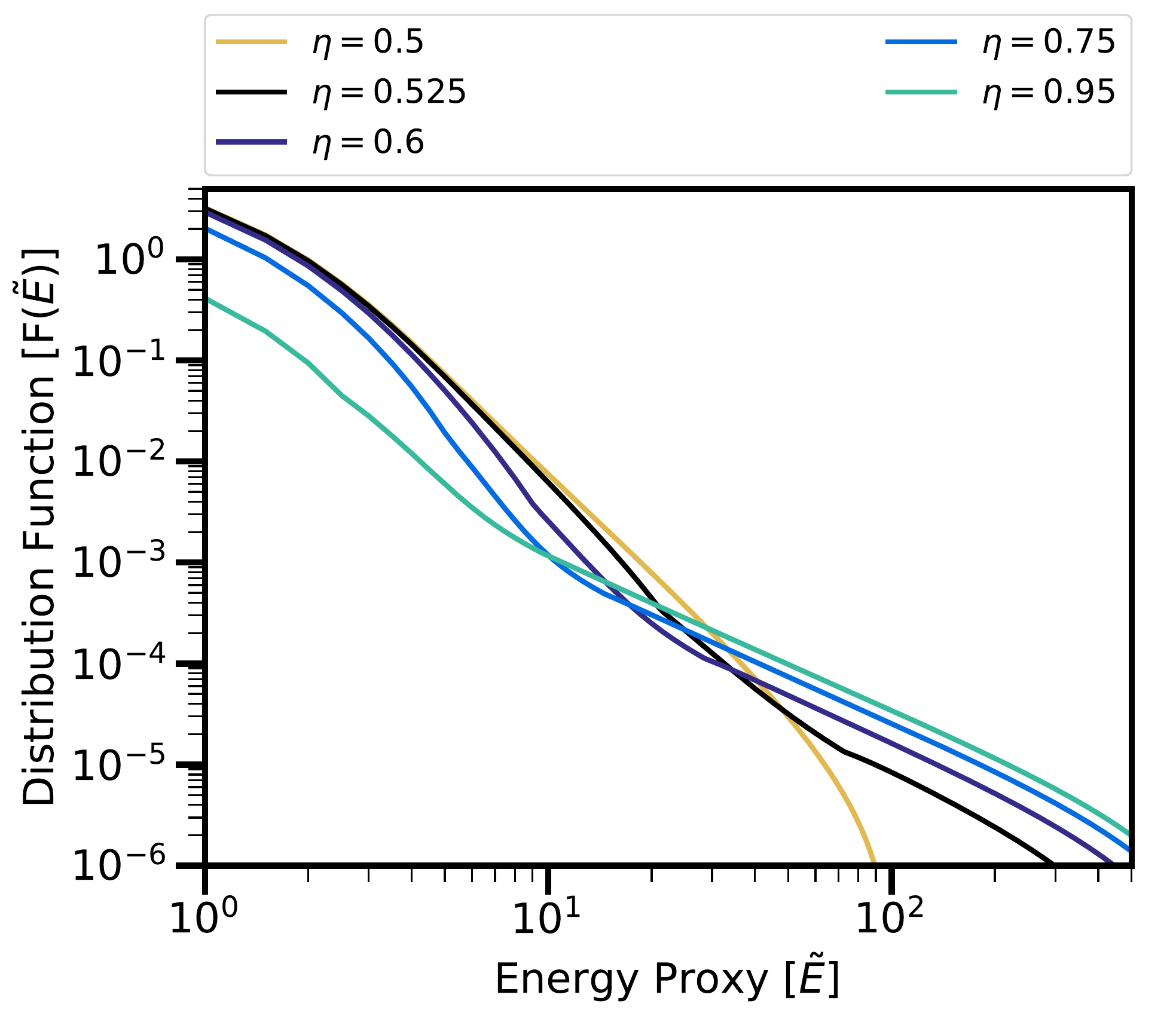}
    \caption{The energy distribution of flaring events for a range of $\eta$. These distributions were computed by performing numerical integrations of Equation \ref{eq:first_energy} for the distributions presented in Figure \ref{fig:positive_negative}. The energy distribution becomes shallower as $\eta$ increases. }\label{fig:energy_distribution}
\end{center}
\end{figure}

\begin{equation}\label{eq:ugly_integral}
\begin{split}
F(E)\sim\int_0^\infty\bigg[\bigg(\frac{w_-}{2\eta-1}\bigg)^{1-\psi}f(w_1)\\\Gamma\bigg(1-\psi,\frac{2\eta-1}{w_-}w_1\bigg)\delta\big(E-4|w_1|C_{\rm crit}\big)\bigg]dw_1+\\
\int_{-\infty}^0\bigg[\bigg(\frac{w_+}{2\eta-1}\bigg)^{1-\beta}f(w_1)\\\Gamma\bigg(1-\beta,\frac{2\eta-1}{w_+}|w_1|\bigg)\delta\big(E-4|w_1|C_{\rm crit}\big)\bigg]dw_1\end{split}
\end{equation}
Using the composition properties of the Dirac delta function, the coefficient  $\delta(E-4|w_1|C_{\rm crit})$ can be written as,
\begin{equation}\label{eq:dirac}
\begin{split}
        \delta(E - 4|w_1|C_{\rm crit}) =\\ \frac{1}{4C_{\rm crit}}\bigg(\delta(w_1 -\frac{ E}{4C_{\rm crit}}) + \delta(w_1 + \frac{ E}{4C_{\rm crit}})\bigg)\,.
\end{split}
\end{equation}
By substituting Equation (\ref{eq:dirac}) into Equation (\ref{eq:ugly_integral}) and performing the integral, Equation (21) in \citet{Berger2009}  generalizes to the following energy distribution function,

\begin{equation}\label{eq:fE_after_integral}
\begin{split}
    F(E)\sim \left(\frac{1}{4C_{\rm crit}}\right)\bigg(\frac{w_-}{2\eta-1}\bigg)^{1-\psi}\\\Gamma\bigg(1-\psi,\frac{2\eta-1}{w_-}\frac{E}{4C_{\rm crit}}\bigg)f\,\bigg(\,\frac{E}{4C_{\rm crit}}\,\bigg)\\+ \left(\frac{1}{4C_{\rm crit}}\right)\bigg(\frac{w_+}{2\eta-1}\bigg)^{1-\beta}\\\Gamma\bigg(1-\beta,\frac{2\eta-1}{w_+}\frac{E}{4C_{\rm crit}}\bigg)f\,\bigg(\,\frac{-E}{4C_{\rm crit}}\,\bigg)\,.
    \end{split}
\end{equation}
By substituting Equation (\ref{eq:f_posneg}) into Equation (\ref{eq:fE_after_integral}), the final energy distribution is given by,
\begin{equation}\label{eq:final_fE}
\begin{split}
    F(E)\sim  \left(\frac{1}{4C_{\rm crit}}\right)\bigg(\frac{w_-}{2\eta-1}\bigg)^{1-\psi}\Gamma\bigg(1-\psi,\frac{2\eta-1}{w_-}\frac{E}{4C_{\rm crit}}\bigg)\\\,\bigg(\,e^{ -(2\eta-1)E/(w_+4C_{\rm crit}) }|\frac{E}{4C_{\rm crit}}|^{-\beta}\,\bigg)\\+\left(\frac{1}{4C_{\rm crit}}\right)\bigg(\frac{w_+}{2\eta-1}\bigg)^{1-\beta}\Gamma\bigg(1-\beta,\frac{2\eta-1}{w_+}\frac{E}{4C_{\rm crit}}\bigg)\\\,\bigg(\,e^{ -(2\eta-1)E/(w_-4C_{\rm crit}) }|\frac{E}{4C_{\rm crit}}|^{-\psi}\,\bigg)\\\,.
    \end{split}
\end{equation}
For large $x$, the incomplete Gamma function can be approximated as the asymptotic series \citep{Temme1975},

\begin{equation}
    \Gamma(a,x)\approx x^{a-1}e^{-x}\bigg(1+(a-1)x^{-1}+(a-1)(a-2)x^{-2}+...\bigg)\,,
\end{equation}
for $x>>|a-n|$, where $n$ is the index of the term number in the series, or for a  series truncated at  the point where $(a-n)/x \sim 1$. Therefore, Equation (\ref{eq:final_fE}) has the form of a sum of power laws times a negative exponential for large flare energies. We verified numerically that  Equation (\ref{eq:final_fE}) has the form of a power law with an exponent $\alpha\approx \beta+\psi-1$ and an exponential cutoff set by $w_-$ (where $w_-<w_+$), analagous to the simpler case found in \citet{Berger2009}, where $\beta=\psi$ and $\alpha= 2\beta-1$. For example, for $w_-=10$, the inertial range extends from $0<w<100$, while for $w_-=1$, it extends from $0<w<10$. As can be seen in Figure \ref{fig:positive_negative}, although the negative exponential dominates for most of the domain for negative winding numbers, $\psi$ slightly decreases as $\eta$ increases. Therefore, as $\eta$ changes,  $\alpha$ scales primarily with $\beta$. 

To validate this analytic calculation, we computed the energy distribution $F(\tilde{E})$ for a  proxy for the flare energy, $\tilde{E}\sim w/4C_{\rm crit}$. We computed this by performing a numerical integration of Equation \ref{eq:first_energy} directly, for the distribution functions presented in Figure \ref{fig:positive_negative}. We show the energy distributions in Figure \ref{fig:energy_distribution}, for a range of $\eta$. As expected, the energy distribution becomes shallower as winding parity violation is increased. 

Our main conclusion is that the slope decreases as $\eta \rightarrow1$. This result can be interpreted physically as follows. As more coherent sequences are injected into the corona, the removal of interchanges yields a larger number of coherent additions of sequences, as shown in Figure~\ref{fig:schematic_braiding_constructive}. Compared to the symmetric case ($\eta=1/2$), this has the effect of allowing larger values of $|w|$ to build up, since the injection of a new braid is now more likely to increase $|w|$ than to decrease it. The winding distribution $f(w)$ thus becomes flatter. Since the flare energy index $\alpha$ is directly related to the winding number index, a flatter distribution of winding numbers immediately produces a flatter energy distribution. This change in slope is compensated for by a reduction in the overall number of flaring events at $\eta\to 1$, so that the total amount of energy dissipated remains the same.

\begin{figure}[]
\begin{center}
    \includegraphics[scale=0.42,angle=0]{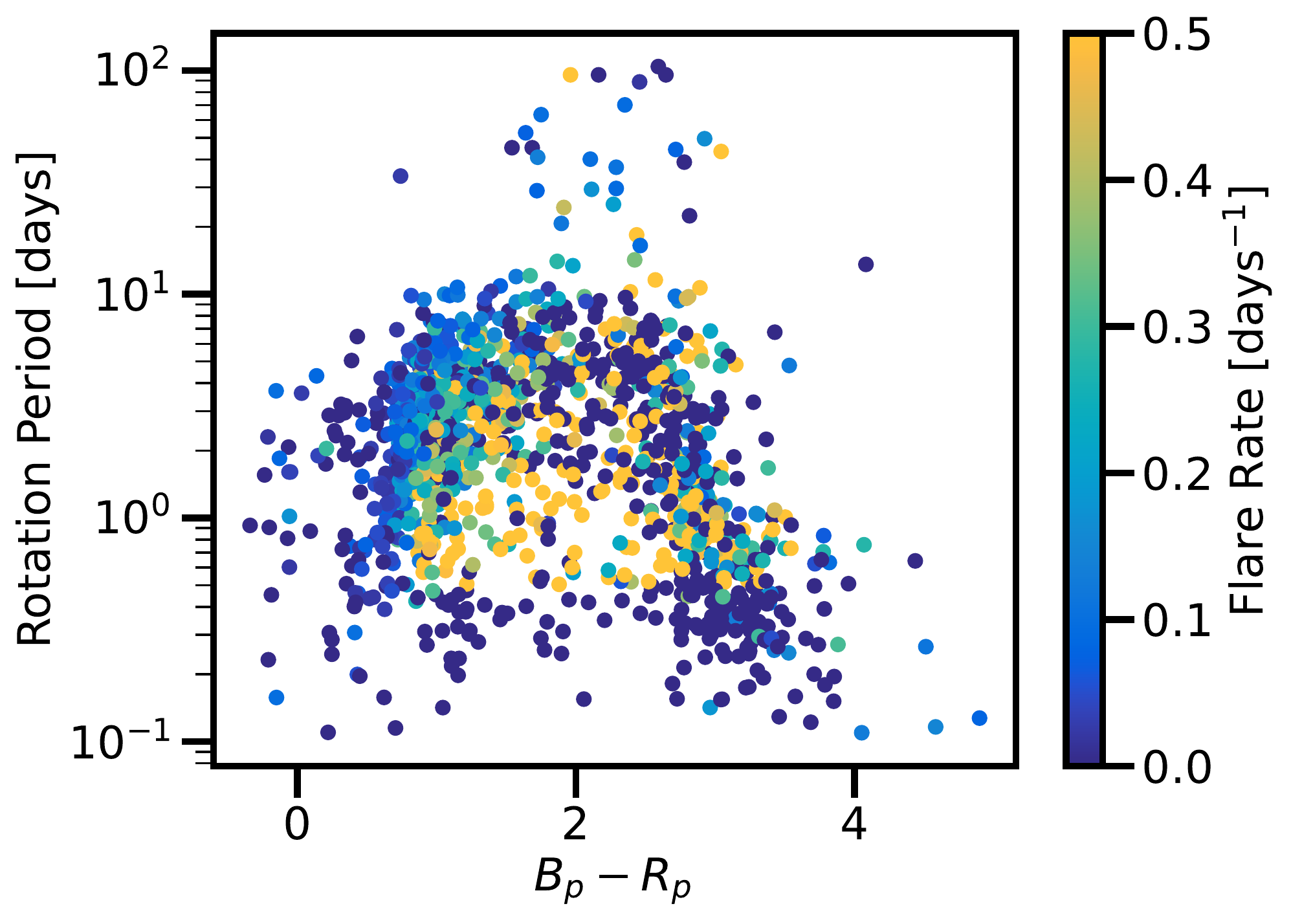}
    \caption{The Gaia DR2 $B_p-R_p$ color of every star in our sample vs. the measured rotation periods from the literature \citep{feinstein20, howard20}. Each point is colored by the flare rate, taken from G{\"u}nther et al. (in prep). All of these stars were observed by \textit{TESS} at 2-minute cadence. Our sample is biased towards stars with rotation periods of $< 13$~days due to the observing strategy of \textit{TESS}.}\label{fig:flares_rotation_rates}
\end{center}
\end{figure}

\section{Connection with TESS Observations }\label{sec:observations}

Studying stellar activity on a statistical level requires the observation and identification of a large number of flaring events. Historically, such a catalogue has been difficult to compile due to the  long observational baselines with high temporal cadences required for each star to capture short-lived  events. It is also a non-trivial task to identify and characterize flares.  Recently, surveys designed to discover extrasolar planets such as \textit{Kepler}/\textit{K2} \citep{Borucki10, Howell2014} and the Transiting Exoplanet Survey Satellite \citep[\textit{TESS};][]{ricker15} have provided a wealth of observations that can be used for compiling a catalogue of flares (see also Section \ref{sec:intro}).

\subsection{Sample \& Flare Identification}

Our calculations suggest that stars with strong Coriolis forces should exhibit a higher proportion of high-energy flares relative to low-energy ones. To verify this empirically, we consider the subset of stars that were observed by \textit{TESS} at 2-minute cadence, are included in our sample from \cite{Feinstein2021}, and which have rotation periods available in the literature. The rotation periods used here were measured from TESS and Evryscope light curves, and were taken directly from \cite{feinstein20} and \cite{howard20}, which yields a sample of 1380 stars ranging from $T_\textrm{eff}=2300-9300$~K. A summary of this sub-sample is presented in Figure~\ref{fig:flares_rotation_rates}. Although requiring a measured period significantly decreases the number of stars in our sample, we still have enough stars to check for signatures of the Coriolis force.

The flares were detected using the convolutional neural network models (CNNs) presented in \cite{feinstein20}. These models were specifically designed to find flares in \textit{TESS} 2-minute cadence light curves and assign a ``probability" to each flare of being real (1) or not (0). We run the 10 CNNs recommended by \cite{feinstein20} and average the prediction outputs for each light curve. The average prediction is used as our probability that each flare is real.

\cite{feinstein20} included all flares with a threshold of $\geq 0.5$ to be potential true flares, meaning the flares in their sample have a 50\% probability of being a true flare. However, we adopt a more conservative threshold of 0.9 to ensure that at least 90\% of the samples are classified as true flares. 

Given the limited size of the training set for the \cite{feinstein20} CNNs, we applied additional false-positive filters from G{\"u}nther et al. (in prep), which are summarized in \cite{Feinstein2021}. The filters remove flares with peak intensities that are within $3\times$root-mean-square of the light curve, flares with equivalent durations of $\leq 4$~minutes, and flares that are associated with periodic events (e.g., eclipsing binaries or variable stars). Once these filters are applied, we are left with a sample of 16,184 flares with true-flare probabilities $\geq 0.9$ from 869 stars.

\subsection{Observational Results}\label{sec:obs_results}

\begin{figure}[]
\begin{center}
    \includegraphics[width=0.47\textwidth, angle=0]{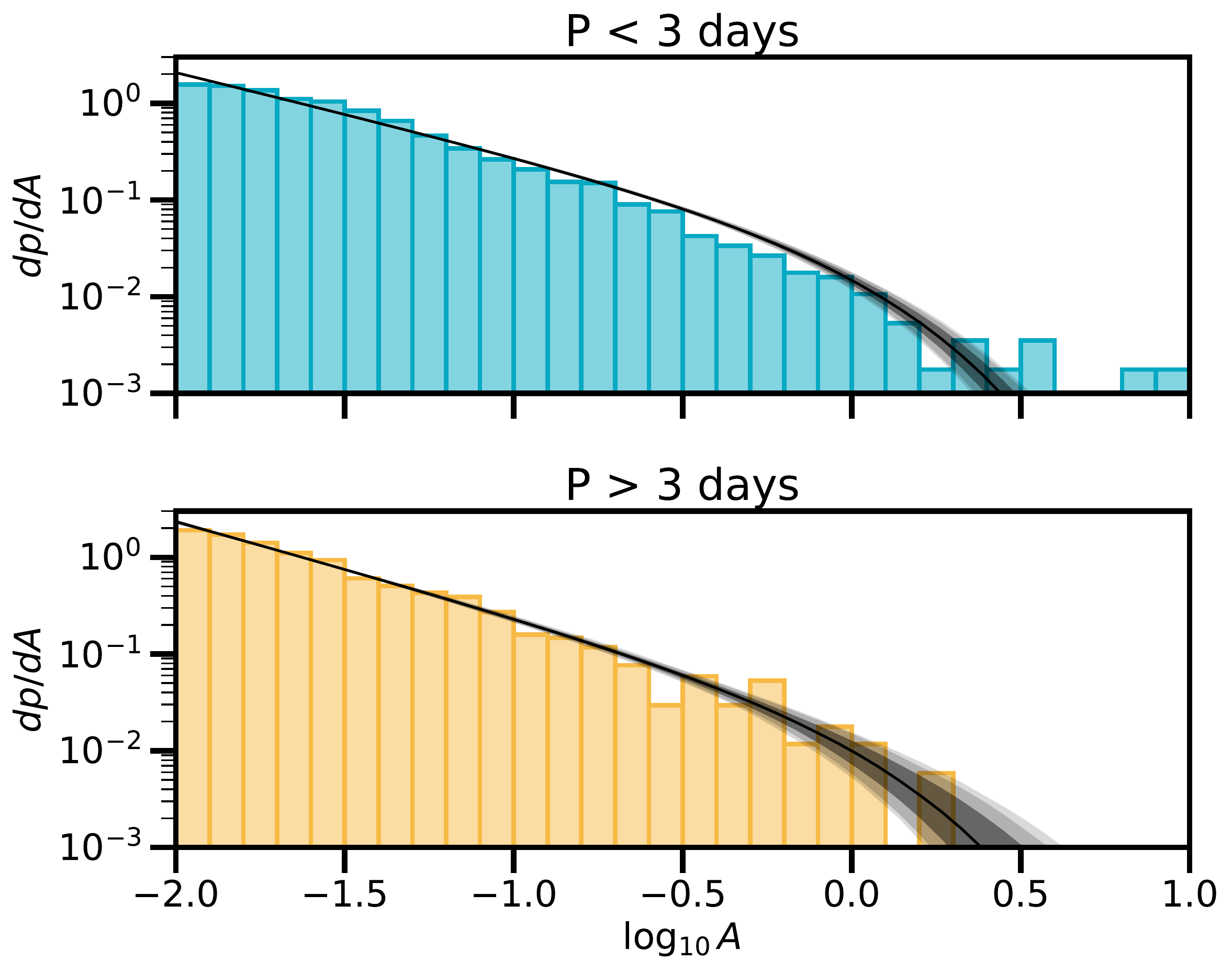}
    \caption{Flare frequency distributions (FFDs) for the flares in our sample, binned by the intensity of the flare in percentage of the star's normalized flux. The flare amplitude, $A$, is divided into bins of equal width in log-space (with 5 bins per dex) ranging from $-2.0 \leq log_{10}(A) \leq 1$, and the vertical axis plots the number of flares observed per star per day in each amplitude bin. The top panel shows the distribution of flares on stars with rotation periods, $P_{\rm rot}$, of $<3$~days (522 stars with 11,614~flares total), and the bottom shows the distribution of flares on stars with $P_{\rm rot} \geq 3$~days (347 stars with 4,570~flares total). While our sample of slower rotators is incomplete, there is evidence that the more rapidly rotating objects have shallower slopes characterized by more energetic events, which  implies stronger winding parity violation. }\label{fig:flares_intensity_distributions}
\end{center}
\end{figure}

\begin{figure}[]
\begin{center}
    \includegraphics[scale=0.235, angle=0]{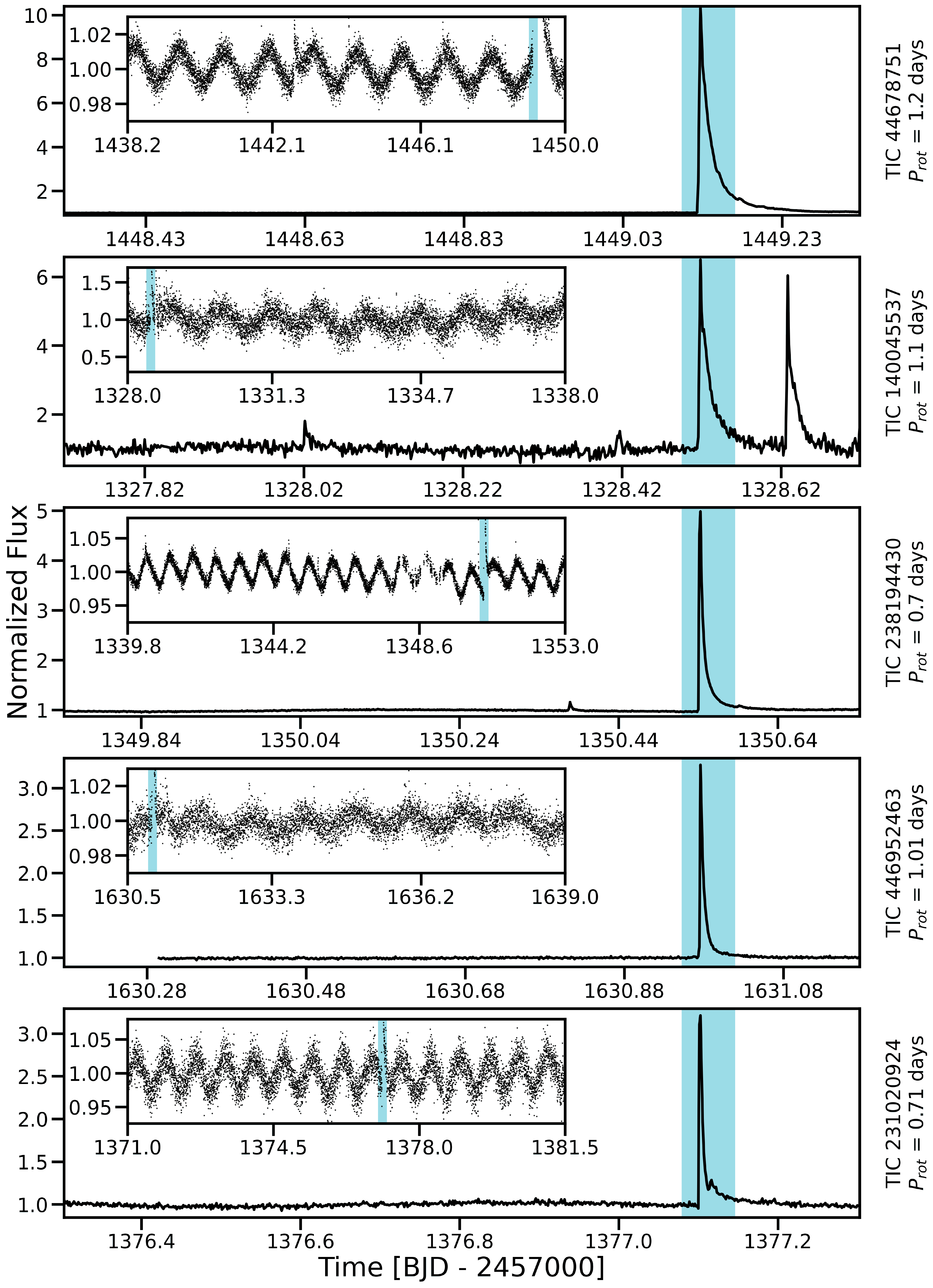}
    \caption{Five of the highest amplitude flares (highlighted in blue) from five different stars in our sample. The inset panels are of the full light curve, displaying various forms of starspot driven modulation, used to measure the rotation periods of these stars. Each of these stars are fast rotators, and contribute to the top panel of Figure \ref{fig:flares_intensity_distributions}. }\label{fig:flare_light_curves}
\end{center}
\end{figure}

\begin{figure}[]
\begin{center}
    \includegraphics[width=0.46\textwidth, angle=0]{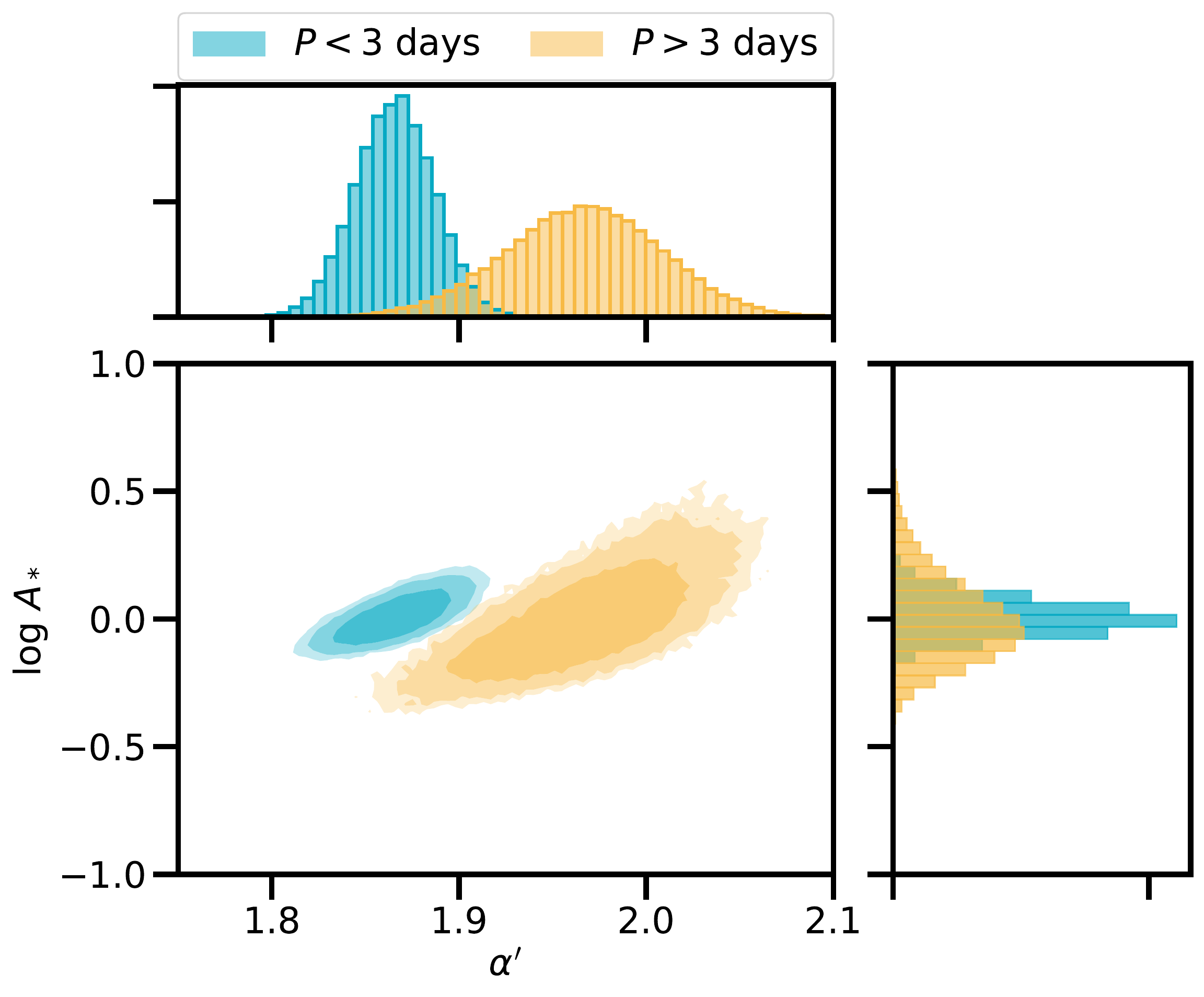}
    \caption{The distributions of the best fitting  power-law exponent and amplitude cutoff parameter for the slow and fast rotators, from the MCMC fitting process described in \S\ref{sec:obs_results}. This shows the full 2D posterior PDF, including both slopes and cutoffs, with histograms showing the marginal 1D PDFs. In the central panel showing the 2D PDF, the shaded regions are the 68\%, 90\%, and 95\% CIs, which visually demonstrate  that the CIs are  overlapping at  68\% confidence. This suggests that the distributions of flares are marginally different for fast and slow rotators. }\label{fig:hist_q}
\end{center} 
\end{figure}

In Figure~\ref{fig:flares_intensity_distributions}, we show the intensity distributions of flaring events for stars that we identify as fast ($P_{\rm rot}< 3$~days) and slow rotators ($P_{\rm rot}\geq 3$~days). In these plots, the flare amplitude, $A$ (expressed as a percentage of the star's normalized flux), is divided into bins of equal width in log-space from $-2.0 \leq log_{10}(A) \leq 1$ (with 5 bins per dex, and 30 bins in total), and the vertical axis shows the number of flares, weighted by their true-flare probability, observed per star per day within each bin. The bins represent the flare rate for all flares with a threshold of $\geq 0.9$. The error bars on the flare rate in each bin are defined as follows. Our upper estimate for the flare rate includes all flares with a threshold $\geq 0.5$. Our lower estimate is for all flares with a threshold $\geq 0.99$. We have visually verified that even the very large amplitude flares we measure are real, and not the result of a failure in the detection algorithm. In Figure \ref{fig:flare_light_curves}, we show the light curves of five of the highest amplitude flares in the fast rotators to reinforce this conclusion.

In order to evaluate the slope of the distribution of flare amplitudes, and test whether there is a significant difference between these two samples, we fit a truncated power-law distribution $dp/dA \propto A^{-\alpha'} e^{-A/A_*}$ for amplitudes $A>A_{\rm min}$,  where $A_{\rm min} = 10^{-2.0}$ is the smallest amplitude flare for which our data are complete -- there is a visible turn-down in the flare frequency distribution below this limit. We perform this fit separately to each of the two sub-samples (fast- and slow-rotators). Note that the form of this equation is similar to Equation \ref{eq:distribution}, although we include the possibility of a truncation at high amplitude, and denote the slope here $\alpha'$ rather than $\alpha$ to indicate that we are fitting the index for flare amplitude, not energy. We fit the slopes using a Markov chain Monte Carlo (MCMC) method implemented with the \texttt{emcee} package \citep{goodman10, Foreman-Mackey13a}, using the log-likelihood function
\begin{equation}
    \log \mathcal{L} = \sum_i \log \left(\mathcal{N} A_i^{-\alpha'}e^{-A/A_*}\right),
\end{equation}
where $A_i$ is the amplitude of the $i$\,th flare in the sample, $A_*$ is a flare amplitude cutoff parameter to be fit, and $\mathcal{N}$ is a normalization factor chosen to ensure that $\int_{A_{\rm min}}^{\infty} (dp/dA)\, dA = 1$. The MCMC fit has two free parameters -- $A_*$ and $\alpha'$ -- and we adopt priors that are flat in $\log A_*$ and $\alpha'$. 

We initialized our MCMC fit with 100 walkers and iterated for 2500 steps; we discard the first 1000 steps for burn-in, and verify visually that the chains are well-converged. We show the median fits obtained by this method, as well as the confidence intervals around them, overlaid on the data in Figures \ref{fig:flares_intensity_distributions}; we show the joint and marginal posterior probability distribution functions (PDFs) for the slope $\alpha'$ and cutoff parameter $A_*$ in Figure \ref{fig:hist_q}. As the plots show, there is a marginally significant difference in the posterior PDFs for the short- and long-period samples — the $2\sigma$ regions do not overlap, but are close, and the marginal parameter values do overlap at the $2\sigma$ level. Repeating the fits using different breaks between fast and slow rotators of 1 day or two days yields qualitatively similar results. Formally, we find $\alpha' = 1.866\pm0.035$ and $\log A_*=0.004_{-0.106}^{+0.131}$ for the short-period sample, and $\alpha' = 1.967_{-0.070}^{+0.068}$ and $\log A_*=-0.017_{-0.223}^{+0.336}$ for the long-period sample; the quoted uncertainties here indicate the 5\textsuperscript{th} to 95\textsuperscript{th} percentile range. To the extent that the differences are real, and not simply statistical noise, we find the fast rotators showing a flatter slope but with a cutoff at somewhat smaller amplitude, and the slow rotators showing a steeper slope but more gradual cutoff. A shallower slope for fast rotators would be consistent with that seen in the literature \citep{hartmann87, maggio87,Doyle2020}, but we emphasise that our present sample does not provide a statistically significant detection.

\begin{figure}[]
\begin{center}
    \includegraphics[width=0.47\textwidth, angle=0]{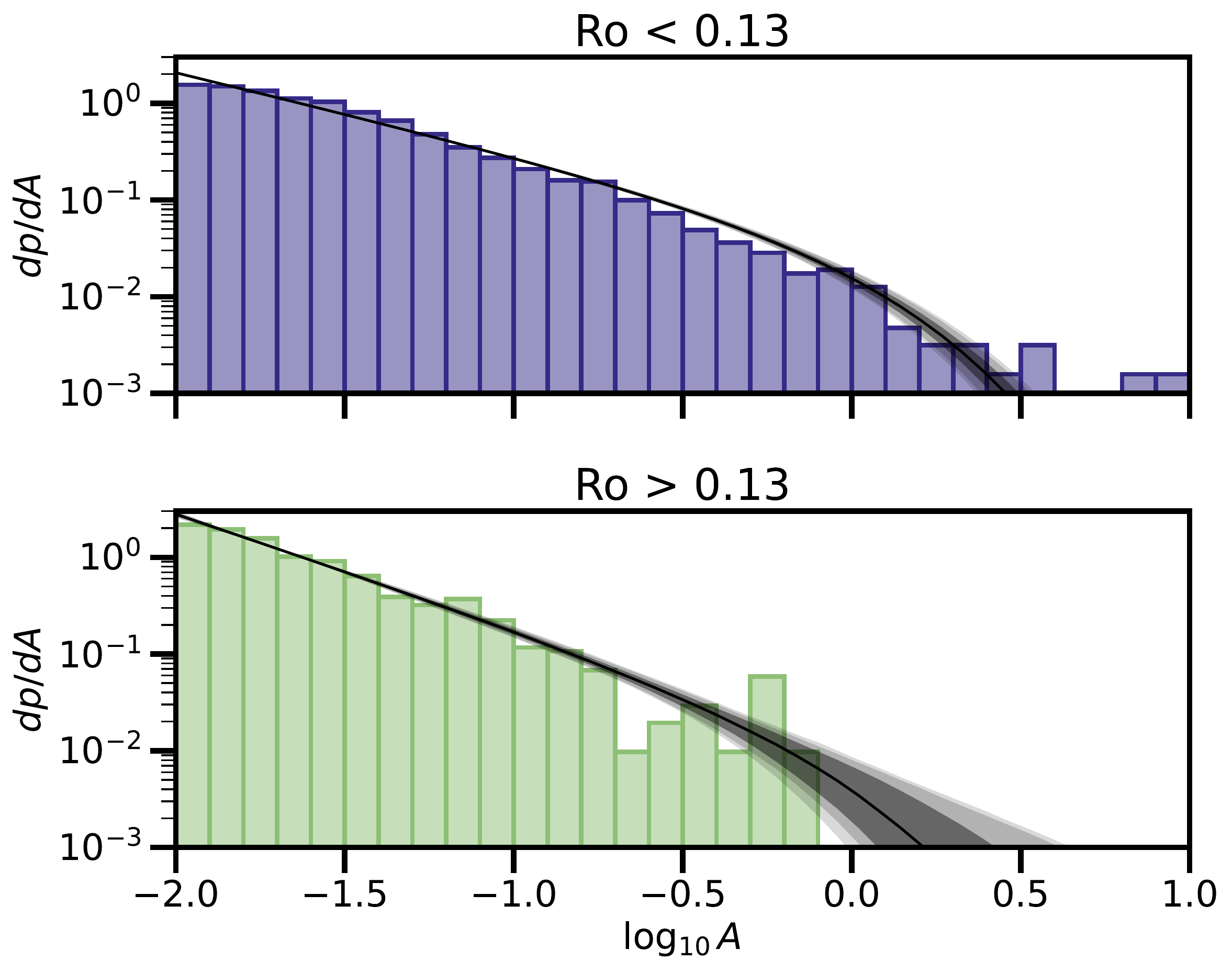}
    \caption{Same as Figure~\ref{fig:flares_intensity_distributions}, with the two sub-samples  grouped by Rossby number. The top panel shows the distribution of flares on stars with Rossby number, $\textrm{Ro}$, of $< 0.13$ (458 stars with 11,1148~flares total), and the bottom panel shows the distribution of flares on stars with $\textrm{Ro} \geq 0.13$ (349 stars with 3,507~flares total). }\label{fig:flares_intensity_distributions_rossby}
\end{center}
\end{figure}

\begin{figure}[]
\begin{center}
    \includegraphics[width=0.47\textwidth, angle=0]{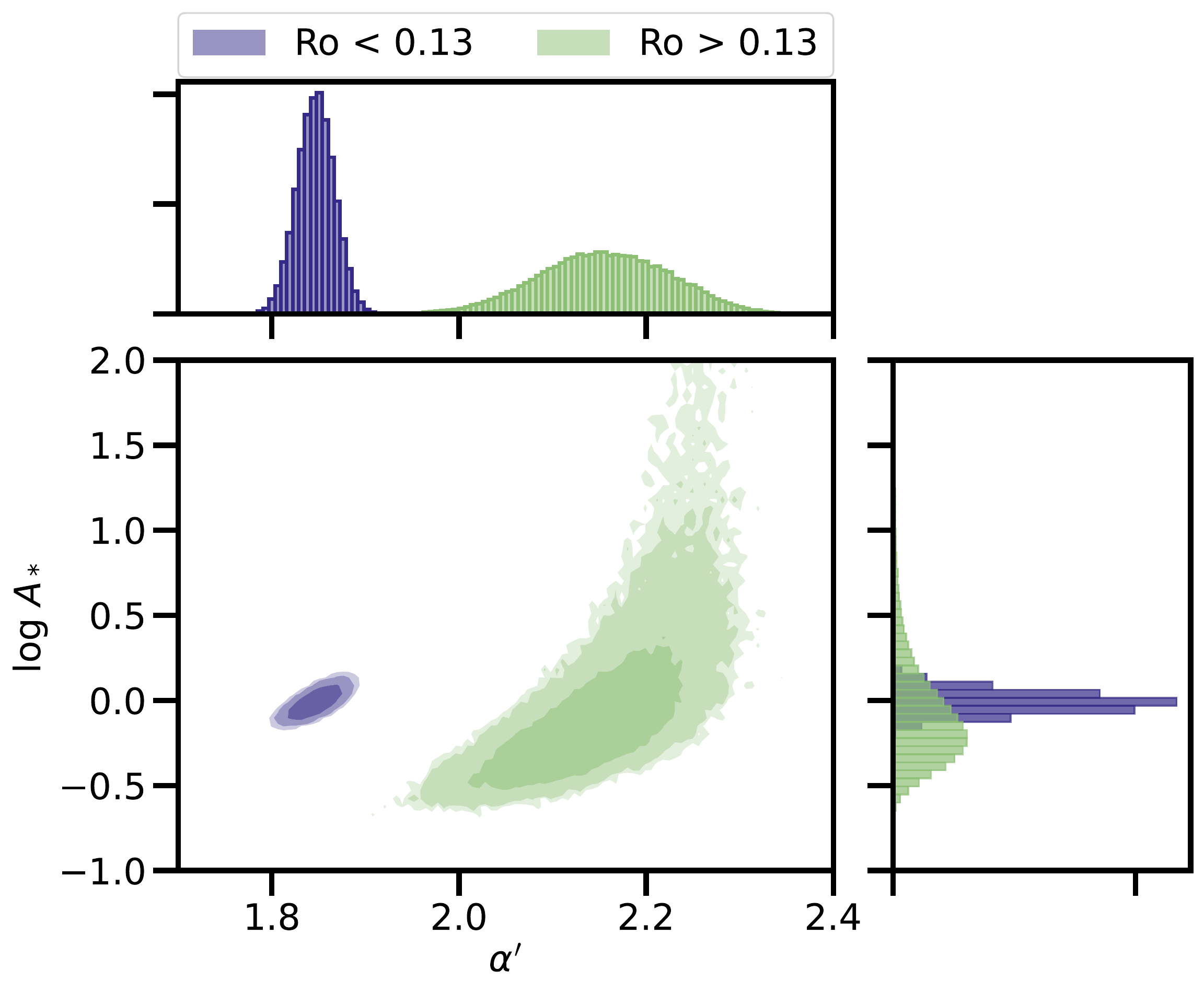}
    \caption{Same as Figure~\ref{fig:hist_q}, with the two sub-samples  grouped by Rossby number. For $\textrm{Ro} < 0.13$, the fits produced a slope $\alpha' = 1.847_{-0.033}^{+0.032}$ and log($A_*$) = $-0.012_{-0.102}^{+0.117}$. For $\textrm{Ro} > 0.13$ the fits produced  $\alpha' = 2.155_{-0.115}^{+0.112}$ and log($A_*$) = $-0.111_{-0.348}^{+1.996}$. }\label{fig:corner_distributions_rossby}
\end{center}
\end{figure}

It has been established that stellar properties other than rotation period, such as mass and surface temperature,  affect  the  dynamo. The Rossby number, $\textrm{Ro}$,  incorporates the  rotational period, mass and temperature, and is defined as $\textrm{Ro}=P/\tau$, where $\tau$ is the convective turnover time \citep{Noyes1984}. For a given stellar type and $\tau$, the Rossby number is a direct proxy for rotational period.   X-ray luminosity and overall activity increases with rotational period for a range of stellar types, including Sun-like stars \citep{Pallavicini1981,wright11,Candelaresi2014} and low mass, fully convective stars \citep{Wright2016}. However, this relationship does not apply for very fast rotators, and the X-ray luminosity saturates for Ro$<0.13$, independant of spectral type \citep{Vilhu1984,Wright2016}.  

Motivated by this work, we repeat the same analysis but group both sub-samples based on Rossby number instead of just the rotational period. We convert stellar mass to the convective turnover time, $\tau$, using Equation 11 in \cite{wright11},

\begin{equation}
    {\rm log}_{10} (\tau) = 1.16–1.49 \,{\rm log}_{10}\bigg(\frac{M}{M_\odot}\bigg) -0.54\, {\rm log}_{10}^2\bigg(\frac{M}{M_\odot}\bigg)\,.
\end{equation}

By binning the stars based on the saturation break of Rossby number, the two resulting sub-samples include 458 stars with 11,1148~flares total where $\textrm{Ro}< 0.13$,  and 349 stars with 3,507~flares total with $\textrm{Ro} \geq 0.13$. The resulting distributions are shown in Figure \ref{fig:flares_intensity_distributions_rossby}, with the best fit distributions overplotted. The fits were calculated in the same  manner as the fits for sub-samples grouped by  rotational period. In Figure \ref{fig:corner_distributions_rossby}, we show the joint and marginal posterior probability distributions functions for slope $\alpha'$ and cutoff parameter $A_*$, with the two sub-samples binned by Rossby number. We find that for $\textrm{Ro} < 0.13$, $\alpha' = 1.847_{-0.033}^{+0.032}$ and log($A_*$) = $-0.012_{-0.102}^{+0.117}$. For $\textrm{Ro} > 0.13$, $\alpha' = 2.155_{-0.115}^{+0.112}$ and log($A_*$) = $-0.111_{-0.348}^{+1.996}$. The results of all fits are presented in Table \ref{table:fits}. It appears that the differences in the distributions are more pronounced when grouped by Rossby number. Since the Rossby number scales with the rotational period, it is plausible that the coriolis effect investigated in this paper is responsible for the differences in the slopes of these FFDs.

There are four potential confounding factors that might influence our results. First and most obviously, the sample is relatively small: only $\approx 3000$ flares in total with amplitude $A>10^{-0.5}$ where we are reasonably complete. It may simply be that the sample is too small to yield statistically-significant results. Second, the measured rotation periods are biased towards $< 12.5$~days due to the observing strategy of \textit{TESS}. Thus our sample of slow rotators is perhaps better described as a sample of moderator rotators; truly slow rotators like the Sun are largely excluded. Third, our sample has  not been corrected for detection biases. Methods such as  injection-recovery tests have been used to adjust observed flare rates in previous studies, as in Figure 11 in \citet{Khovaris2020}. We do not perform injection-recovery tests on this sample for two reasons. For one, \citet{feinstein20} demonstrated that injection recovery tests did not produce accurate results for their CNN, due to the differences in the shape of the flares themselves. Additionally, we only start at 0.32\% amplitudes here and therefore ignore the smallest flares (which are most affected by this bias). A final confounding factor is that, while rotation period is a reasonable proxy for the importance of Coriolis forces in the dynamo, it is certainly not the only factor that might affect flare distributions. In particular, stars of different ages and effective temperature almost certainly have different surface magnetic field strengths, and our fast and slow rotator samples mix together a wide range of stellar age and effective temperature. Similar to our marginal agreement here, \cite{feinstein20} measured the FFD slope for stars with age $\leq 50$~Myr and $> 50$~Myr and found they were in a 1-$\sigma$ agreement with each other. For the specific details of this analysis, we refer the reader to Figures 13-14 and Section 4.1 in \citet{feinstein20}. While the results from our analysis incorporating the Rossby number are highly suggestive,  it is possible that an analysis that separates all of these contributing factors -- age, rotation period, and effective temperature -- might yield stronger results.  

\begin{deluxetable}{l r r}[!ht]\label{table:fits}
\tabletypesize{\footnotesize}
\tablecaption{Summary of the parameter fits for the flare frequency distributions binned by rotation period and Rossby number. \label{tab:stats}}
\tablehead{\colhead{Sample} & \colhead{slope, $\alpha'$} & \colhead{log($A_*$)}  \\
\hline
$P_\textrm{rot} < 3$~days & $1.866 \pm 0.035$ & $0.004_{-0.106}^{+0.131}$\\
$P_\textrm{rot} \geq 3$~days & $1.967_{-0.070}^{+0.068}$ & $-0.017_{-0.223}^{+0.336}$\\
\hline
\hline
$\textrm{Ro} < 0.13$ & $1.847_{-0.033}^{+0.032}$ & $-0.012_{-0.102}^{+0.117}$\\
$\textrm{Ro} \geq 0.13$ & $2.155_{-0.115}^{+0.112}$ & $-0.111_{-0.348}^{+1.996}$
}
\startdata
\enddata
\end{deluxetable}

\section{Conclusions}\label{sec:conclusions}

In this paper, we have investigated the role of the Coriolis force in stellar dynamos.  More specifically, we considered the direct current model of coronal heating \citep{Parker1972} along with the model of magnetic braiding and  reconnection (proposed  by \citealt{Berger2009}; see also \citealt{Berger2015}). We expanded the braiding model to incorporate the effects of the Coriolis force, which should be more dominant in faster rotating stars. Specifically, we incorporated  a bias in handedness of injected  braids, which should be present in stars with dynamos that are strongly affected by the Coriolis force. As increasingly coherent braids are injected into the corona, the slopes of the resulting power-law distributions of energetic flaring events decreases in magnitude (corresponding to a weaker decline in occurrence rates toward increasing flare energies). We search for this effect in the flare frequency distributions for stars observed by \textit{TESS} that have measured rotation periods; however, while the results are suggestively consistent with the theoretical prediction, the sample is too small to yield a definitive statistical conclusion.

While the results presented in this study are suggestive, there are many opportunities for future theoretical work. It would be informative to perform numerical experiments of reconnection events for both the DC and AC regimes of coronal heating. Specifically, numerical simulations of avalanching reconnection events similar to those presented by \citet{lu1991} may also exhibit shallower slopes with driving that mimics effects of the Coriolis force. Being able to directly probe the relation between the magnitude of the Coriolis force and the $\eta$ (handedness) parameter would be an invaluable measurement to test the role of rotation on the topology of the flux tubes and hence the energy of the flaring events. This type of calculation would also yield insights into the connection between the theory of self-organized criticality and the DC braiding and reconnection picture of coronal heating.

Alternatively, it is possible that the power-law distributions of stellar flares are simply realizations of forced MHD turbulence, along the framework of the AC regime. Therefore, numerical simulations of forced coronal plasmas that track reconnection events may reveal an alternative explanation for the differences in the observed slopes.

Several possible adjustments to the observational data could further validate the hypothesis presented here. We have included a relatively small ($\approx 3,000$~stars) sample of stars with measured rotation periods. However, there are $10^5$~stars with high-cadence observations from \textit{TESS}. By measuring the rotation periods for more stars with known flaring events, we would be able to vastly increase the statistical sample in this study. This analysis would also be improved by the proposed increase from a 10- to 3-minute cadence of \textit{TESS} Full-Frame Images for the 2\textsuperscript{nd} extended mission. This improvement would increase our sample of stars by an order of magnitude, which would improve the slope measurements.

It will also be necessary to develop a reliable method for measuring long rotation periods ($P> 13$~days) in \textit{TESS} \citep[e.g.][]{lu2020, breton21, claytor21}. In the current state, we are limited both in the baseline for the \textit{TESS} sector observing strategy ($\sim 27$~days) and the orbital gap halfway through each sector. Traditional methods of measuring rotation periods often identify the systematics that are associated with the beginning or end of each orbit. We are thus limited to only reliably being able to measure short rotation periods.  Alternatively, indirect estimates of stellar rotation periods may be useful for expanding our sample \citep{Mamajek2008}.

It would also be useful to improve the convolutional neural networks presented in \cite{feinstein20} to detect flares on rapid rotating stars ($P< 1$~day). When these methods were developed, there were relatively few examples of these light curves, so that the original models were trained on a limited number of high energy events on rapidly rotating stars. As a result, the sharp rotational features are often confused for flare events. Now that more high energy events have been observed, we have the opportunity to improve our method for flare identification using machine learning. 

Finally, it would be of significant interest to test the impact of stellar age versus rotation period on the flare rates. We were not able to do this analysis with our relatively small sample. However, populations within young stellar clusters, such as Pleiades, Hyades, or Praesepe, have well constrained masses and ages. The activity in young clusters has already been examined using \textit{K2} long-cadence data \citep{ilin19,ilin21}, although this cadence is inadequate for detecting low-amplitude flares. Examining activity in young stellar clusters using new high-cadence \textit{TESS} data may yield key insights into the relationship between rotation, dynamo, and spin-down.


\section{Acknowledgements}
We thank Mitch Berger, Mahboubeh Asgari-Targhi, Eugene Chiang, Amir Siraj, Jared Siegel,  Dimitar Sasselov, Konstantin Batygin, Jacob Bean, Juliette Becker, Fred Ciesla, and Harriet Lau for useful conversations. We thank the two anonymous reviewers for insightful comments and constructive suggestions that strengthened the scientific content of this manuscript, especially for suggesting searching for differences in the power law distribution based on the Rossby number.

ADF acknowledges support from the National Science Foundation Graduate Research Fellowship Program under Grant No. DGE-1746045. JRB~acknowledges financial support from the Australian National University, via the Deakin PhD and Dean's Higher Degree Research (theoretical physics) Scholarships and the Australian Government via the Australian Government Research Training Program Fee-Offset Scholarship. MRK acknowledges support from the Australian Research Council through its \textit{Future Fellowships} scheme, award FT180100375. LAR gratefully acknowledges support from the Research Corporation for Science Advancement through a Cottrell Scholar Award.
C.~F.~acknowledges funding provided by the Australian Research Council (Future Fellowship FT180100495), and the Australia-Germany Joint Research Cooperation Scheme (UA-DAAD). MF is grateful for the support he received from the Hauck Foundation.


\facilities{\textit{TESS} \citep{ricker15}, \textit{Kepler} \citep{Borucki10,Howell2014}}

\software{\texttt{lightkurve} \citep{lightkurve},
          \texttt{matplotlib} \citep{matplotlib},
          \texttt{astropy} \citep{astropy:2013, astropy18},
          \texttt{astroquery}\citep{astroquery19},
          \texttt{numpy} \citep{numpy,numpy2020},
          \texttt{stella} \citep{feinstein20_joss},
          \texttt{emcee} \citep{Foreman-Mackey13a}
          }

\bibliography{sample631}{}

\begin{thebibliography}{}
\expandafter\ifx\csname natexlab\endcsname\relax\def\natexlab#1{#1}\fi
\providecommand{\url}[1]{\href{#1}{#1}}
\providecommand{\dodoi}[1]{doi:~\href{http://doi.org/#1}{\nolinkurl{#1}}}
\providecommand{\doeprint}[1]{\href{http://ascl.net/#1}{\nolinkurl{http://ascl.net/#1}}}
\providecommand{\doarXiv}[1]{\href{https://arxiv.org/abs/#1}{\nolinkurl{https://arxiv.org/abs/#1}}}

\bibitem[{{Abramenko} {et~al.}(1997){Abramenko}, {Wang}, \&
  {Yurchishin}}]{Abramenko1997}
{Abramenko}, V.~I., {Wang}, T., \& {Yurchishin}, V.~B. 1997, \solphys, 174,
  291, \dodoi{10.1023/A:1004957515498}

\bibitem[{{Andrade} {et~al.}(1998){Andrade}, {Schellnhuber}, \&
  {Claussen}}]{Andrade1998}
{Andrade}, R.~F.~S., {Schellnhuber}, H.~J., \& {Claussen}, M. 1998, Physica A
  Statistical Mechanics and its Applications, 254, 557,
  \dodoi{10.1016/S0378-4371(98)00057-0}

\bibitem[{{Antolin} {et~al.}(2021){Antolin}, {Pagano}, {Testa}, {Petralia}, \&
  {Reale}}]{Antolin2021}
{Antolin}, P., {Pagano}, P., {Testa}, P., {Petralia}, A., \& {Reale}, F. 2021,
  Nature Astronomy, 5, 54, \dodoi{10.1038/s41550-020-1199-8}

\bibitem[{{Aschwanden} {et~al.}(1998){Aschwanden}, {Dennis}, \&
  {Benz}}]{Aschwanden1998}
{Aschwanden}, M.~J., {Dennis}, B.~R., \& {Benz}, A.~O. 1998, \apj, 497, 972,
  \dodoi{10.1086/305484}

\bibitem[{{Aschwanden} \& {G{\"u}del}(2021)}]{Aschwanden21}
{Aschwanden}, M.~J., \& {G{\"u}del}, M. 2021, The Astrophysical Journal, 910,
  41, \dodoi{10.3847/1538-4357/abdec7}

\bibitem[{{Asgari-Targhi} {et~al.}(2015){Asgari-Targhi}, {Schmelz}, {Imada},
  {Pathak}, \& {Christian}}]{Asgari2015}
{Asgari-Targhi}, M., {Schmelz}, J.~T., {Imada}, S., {Pathak}, S., \&
  {Christian}, G.~M. 2015, \apj, 807, 146, \dodoi{10.1088/0004-637X/807/2/146}

\bibitem[{{Asgari-Targhi} \& {van Ballegooijen}(2012)}]{Asgari2012}
{Asgari-Targhi}, M., \& {van Ballegooijen}, A.~A. 2012, \apj, 746, 81,
  \dodoi{10.1088/0004-637X/746/1/81}

\bibitem[{{Asgari-Targhi} {et~al.}(2013){Asgari-Targhi}, {van Ballegooijen},
  {Cranmer}, \& {DeLuca}}]{Asgari2013}
{Asgari-Targhi}, M., {van Ballegooijen}, A.~A., {Cranmer}, S.~R., \& {DeLuca},
  E.~E. 2013, \apj, 773, 111, \dodoi{10.1088/0004-637X/773/2/111}

\bibitem[{{Asgari-Targhi} {et~al.}(2014){Asgari-Targhi}, {van Ballegooijen}, \&
  {Imada}}]{Asgari2014}
{Asgari-Targhi}, M., {van Ballegooijen}, A.~A., \& {Imada}, S. 2014, \apj, 786,
  28, \dodoi{10.1088/0004-637X/786/1/28}

\bibitem[{{Astropy Collaboration} {et~al.}(2013){Astropy Collaboration},
  {Robitaille}, {Tollerud}, {Greenfield}, {Droettboom}, {Bray}, {Aldcroft},
  {Davis}, {Ginsburg}, {Price-Whelan}, {Kerzendorf}, {Conley}, {Crighton},
  {Barbary}, {Muna}, {Ferguson}, {Grollier}, {Parikh}, {Nair}, {Unther},
  {Deil}, {Woillez}, {Conseil}, {Kramer}, {Turner}, {Singer}, {Fox}, {Weaver},
  {Zabalza}, {Edwards}, {Azalee Bostroem}, {Burke}, {Casey}, {Crawford},
  {Dencheva}, {Ely}, {Jenness}, {Labrie}, {Lim}, {Pierfederici}, {Pontzen},
  {Ptak}, {Refsdal}, {Servillat}, \& {Streicher}}]{astropy:2013}
{Astropy Collaboration}, {Robitaille}, T.~P., {Tollerud}, E.~J., {et~al.} 2013,
  \aap, 558, A33, \dodoi{10.1051/0004-6361/201322068}

\bibitem[{{Audard} {et~al.}(2000){Audard}, {G{\"u}del}, {Drake}, \&
  {Kashyap}}]{Audard2000}
{Audard}, M., {G{\"u}del}, M., {Drake}, J.~J., \& {Kashyap}, V.~L. 2000, \apj,
  541, 396, \dodoi{10.1086/309426}

\bibitem[{{Babcock} \& {Westervelt}(1990)}]{Babcock1990}
{Babcock}, K.~L., \& {Westervelt}, R.~M. 1990, \prl, 64, 2168,
  \dodoi{10.1103/PhysRevLett.64.2168}

\bibitem[{{Bak} {et~al.}(1989){Bak}, {Chen}, \& {Creutz}}]{bak1989}
{Bak}, P., {Chen}, K., \& {Creutz}, M. 1989, \nat, 342, 780,
  \dodoi{10.1038/342780a0}

\bibitem[{{Bak} {et~al.}(1990){Bak}, {Chen}, \& {Tang}}]{Bak1990}
{Bak}, P., {Chen}, K., \& {Tang}, C. 1990, Physics Letters A, 147, 297,
  \dodoi{10.1016/0375-9601(90)90451-S}

\bibitem[{{Bak} {et~al.}(1997){Bak}, {Paczuski}, \& {Shubik}}]{Bak1997}
{Bak}, P., {Paczuski}, M., \& {Shubik}, M. 1997, Physica A Statistical
  Mechanics and its Applications, 246, 430,
  \dodoi{10.1016/S0378-4371(97)00401-9}

\bibitem[{{Bak} \& {Sneppen}(1993)}]{Bak1993}
{Bak}, P., \& {Sneppen}, K. 1993, \prl, 71, 4083,
  \dodoi{10.1103/PhysRevLett.71.4083}

\bibitem[{{Bak} \& {Tang}(1989)}]{bak1989_earthquakes}
{Bak}, P., \& {Tang}, C. 1989, \jgr, 94, 15,635,
  \dodoi{10.1029/JB094iB11p15635}

\bibitem[{{Bak} {et~al.}(1987){Bak}, {Tang}, \& {Wiesenfeld}}]{bak1987}
{Bak}, P., {Tang}, C., \& {Wiesenfeld}, K. 1987, \prl, 59, 381,
  \dodoi{10.1103/PhysRevLett.59.381}

\bibitem[{{Bak} {et~al.}(1988){Bak}, {Tang}, \& {Wiesenfeld}}]{bak1988}
---. 1988, \pra, 38, 364, \dodoi{10.1103/PhysRevA.38.364}

\bibitem[{{Bao} \& {Zhang}(1998)}]{Bao1998}
{Bao}, S., \& {Zhang}, H. 1998, \apjl, 496, L43, \dodoi{10.1086/311232}

\bibitem[{{Berger}(1993)}]{Berger1993}
{Berger}, M.~A. 1993, \prl, 70, 705, \dodoi{10.1103/PhysRevLett.70.705}

\bibitem[{{Berger} \& {Asgari-Targhi}(2009)}]{Berger2009}
{Berger}, M.~A., \& {Asgari-Targhi}, M. 2009, \apj, 705, 347,
  \dodoi{10.1088/0004-637X/705/1/347}

\bibitem[{{Berger} {et~al.}(2015){Berger}, {Asgari-Targhi}, \&
  {Deluca}}]{Berger2015}
{Berger}, M.~A., {Asgari-Targhi}, M., \& {Deluca}, E.~E. 2015, Journal of
  Plasma Physics, 81, 395810404, \dodoi{10.1017/S0022377815000483}

\bibitem[{{Borucki} {et~al.}(2010){Borucki}, {Koch}, {Basri}, {Batalha},
  {Brown}, {Caldwell}, {Caldwell}, {Christensen-Dalsgaard}, {Cochran},
  {DeVore}, {Dunham}, {Dupree}, {Gautier}, {Geary}, {Gilliland}, {Gould},
  {Howell}, {Jenkins}, {Kondo}, {Latham}, {Marcy}, {Meibom}, {Kjeldsen},
  {Lissauer}, {Monet}, {Morrison}, {Sasselov}, {Tarter}, {Boss}, {Brownlee},
  {Owen}, {Buzasi}, {Charbonneau}, {Doyle}, {Fortney}, {Ford}, {Holman},
  {Seager}, {Steffen}, {Welsh}, {Rowe}, {Anderson}, {Buchhave}, {Ciardi},
  {Walkowicz}, {Sherry}, {Horch}, {Isaacson}, {Everett}, {Fischer}, {Torres},
  {Johnson}, {Endl}, {MacQueen}, {Bryson}, {Dotson}, {Haas}, {Kolodziejczak},
  {Van Cleve}, {Chandrasekaran}, {Twicken}, {Quintana}, {Clarke}, {Allen},
  {Li}, {Wu}, {Tenenbaum}, {Verner}, {Bruhweiler}, {Barnes}, \&
  {Prsa}}]{Borucki10}
{Borucki}, W.~J., {Koch}, D., {Basri}, G., {et~al.} 2010, Science, 327, 977,
  \dodoi{10.1126/science.1185402}

\bibitem[{{Breton} {et~al.}(2021){Breton}, {Santos}, {Bugnet}, {Mathur},
  {Garc{\'\i}a}, \& {Pall{\'e}}}]{breton21}
{Breton}, S.~N., {Santos}, A.~R.~G., {Bugnet}, L., {et~al.} 2021, \aap, 647,
  A125, \dodoi{10.1051/0004-6361/202039947}

\bibitem[{{Brown} {et~al.}(1998){Brown}, {McArthur}, {Barrett}, {McIntosh}, \&
  {Emslie}}]{Brown1998}
{Brown}, J.~C., {McArthur}, G.~K., {Barrett}, R.~K., {McIntosh}, S.~W., \&
  {Emslie}, A.~G. 1998, \solphys, 179, 379, \dodoi{10.1023/A:1005011107402}

\bibitem[{{Browning} {et~al.}(2016){Browning}, {Weber}, {Chabrier}, \&
  {Massey}}]{Browning2016}
{Browning}, M.~K., {Weber}, M.~A., {Chabrier}, G., \& {Massey}, A.~P. 2016,
  \apj, 818, 189, \dodoi{10.3847/0004-637X/818/2/189}

\bibitem[{{Candelaresi} {et~al.}(2014){Candelaresi}, {Hillier}, {Maehara},
  {Brandenburg}, \& {Shibata}}]{Candelaresi2014}
{Candelaresi}, S., {Hillier}, A., {Maehara}, H., {Brandenburg}, A., \&
  {Shibata}, K. 2014, \apj, 792, 67, \dodoi{10.1088/0004-637X/792/1/67}

\bibitem[{{Cargill}(1994)}]{Cargill1994}
{Cargill}, P.~J. 1994, \apj, 422, 381, \dodoi{10.1086/173733}

\bibitem[{{Carlson} \& {Langer}(1989)}]{Carlson1989}
{Carlson}, J.~M., \& {Langer}, J.~S. 1989, \prl, 62, 2632,
  \dodoi{10.1103/PhysRevLett.62.2632}

\bibitem[{{Carrington}(1859)}]{Carrington1859}
{Carrington}, R.~C. 1859, \mnras, 20, 13, \dodoi{10.1093/mnras/20.1.13}

\bibitem[{{Charbonneau}(2010)}]{Charbonneau2010}
{Charbonneau}, P. 2010, Living Reviews in Solar Physics, 7, 3,
  \dodoi{10.12942/lrsp-2010-3}

\bibitem[{{Charbonneau} {et~al.}(2001){Charbonneau}, {McIntosh}, {Liu}, \&
  {Bogdan}}]{Charbonneau2001}
{Charbonneau}, P., {McIntosh}, S.~W., {Liu}, H.-L., \& {Bogdan}, T.~J. 2001,
  \solphys, 203, 321, \dodoi{10.1023/A:1013301521745}

\bibitem[{{Choudhuri} \& {D'Silva}(1990)}]{Choudhuri1990}
{Choudhuri}, A.~R., \& {D'Silva}, S. 1990, \aap, 239, 326

\bibitem[{{Choudhuri} \& {Gilman}(1987)}]{Choudhuri1987}
{Choudhuri}, A.~R., \& {Gilman}, P.~A. 1987, \apj, 316, 788,
  \dodoi{10.1086/165243}

\bibitem[{{Claytor} {et~al.}(2021){Claytor}, {van Saders}, {Llama}, {Sadowski},
  {Quach}, \& {Avallone}}]{claytor21}
{Claytor}, Z.~R., {van Saders}, J.~L., {Llama}, J., {et~al.} 2021, arXiv
  e-prints, arXiv:2104.14566.
\newblock \doarXiv{2104.14566}

\bibitem[{{Crosby} {et~al.}(1993){Crosby}, {Aschwanden}, \&
  {Dennis}}]{Crosby1993}
{Crosby}, N.~B., {Aschwanden}, M.~J., \& {Dennis}, B.~R. 1993, \solphys, 143,
  275, \dodoi{10.1007/BF00646488}

\bibitem[{{Dahlburg} {et~al.}(2005){Dahlburg}, {Klimchuk}, \&
  {Antiochos}}]{Dahlburg2005}
{Dahlburg}, R.~B., {Klimchuk}, J.~A., \& {Antiochos}, S.~K. 2005, \apj, 622,
  1191, \dodoi{10.1086/425645}

\bibitem[{{Datlowe} {et~al.}(1974){Datlowe}, {Elcan}, \&
  {Hudson}}]{Datlowe1974}
{Datlowe}, D.~W., {Elcan}, M.~J., \& {Hudson}, H.~S. 1974, \solphys, 39, 155,
  \dodoi{10.1007/BF00154978}

\bibitem[{{Davenport}(2016)}]{Davenport16}
{Davenport}, J.~R.~A. 2016, The Astrophysical Journal, 829, 23,
  \dodoi{10.3847/0004-637X/829/1/23}

\bibitem[{{de Arcangelis} {et~al.}(2006){de Arcangelis}, {Godano}, {Lippiello},
  \& {Nicodemi}}]{deArcangelis06}
{de Arcangelis}, L., {Godano}, C., {Lippiello}, E., \& {Nicodemi}, M. 2006,
  \prl, 96, 051102, \dodoi{10.1103/PhysRevLett.96.051102}

\bibitem[{{Dendy} {et~al.}(1998){Dendy}, {Helander}, \& {Tagger}}]{Dendy1998}
{Dendy}, R.~O., {Helander}, P., \& {Tagger}, M. 1998, \aap, 337, 962.
\newblock \doarXiv{astro-ph/9907055}

\bibitem[{{Dennis}(1985)}]{Dennis1985}
{Dennis}, B.~R. 1985, \solphys, 100, 465, \dodoi{10.1007/BF00158441}

\bibitem[{{Dmitruk} \& {G{\'o}mez}(1997)}]{Dmitruk1997}
{Dmitruk}, P., \& {G{\'o}mez}, D.~O. 1997, \apjl, 484, L83,
  \dodoi{10.1086/310760}

\bibitem[{{Doyle} {et~al.}(2020){Doyle}, {Ramsay}, \& {Doyle}}]{Doyle2020}
{Doyle}, L., {Ramsay}, G., \& {Doyle}, J.~G. 2020, \mnras, 494, 3596,
  \dodoi{10.1093/mnras/staa923}

\bibitem[{{Doyle} {et~al.}(2019){Doyle}, {Ramsay}, {Doyle}, \&
  {Wu}}]{Doyle2019}
{Doyle}, L., {Ramsay}, G., {Doyle}, J.~G., \& {Wu}, K. 2019, \mnras, 489, 437,
  \dodoi{10.1093/mnras/stz2205}

\bibitem[{{Doyle} {et~al.}(2018){Doyle}, {Ramsay}, {Doyle}, {Wu}, \&
  {Scullion}}]{Doyle2018}
{Doyle}, L., {Ramsay}, G., {Doyle}, J.~G., {Wu}, K., \& {Scullion}, E. 2018,
  \mnras, 480, 2153, \dodoi{10.1093/mnras/sty1963}

\bibitem[{{Drake}(1971)}]{Drake1971}
{Drake}, J.~F. 1971, \solphys, 16, 152, \dodoi{10.1007/BF00154510}

\bibitem[{{D'Silva} \& {Choudhuri}(1993)}]{DSilva1993}
{D'Silva}, S., \& {Choudhuri}, A.~R. 1993, \aap, 272, 621

\bibitem[{{D'Silva} \& {Howard}(1995)}]{DSilva1995}
{D'Silva}, S., \& {Howard}, R.~F. 1995, \solphys, 159, 63,
  \dodoi{10.1007/BF00733032}

\bibitem[{{Einaudi} \& {Velli}(1999)}]{Einaudi1999}
{Einaudi}, G., \& {Velli}, M. 1999, Physics of Plasmas, 6, 4146,
  \dodoi{10.1063/1.873679}

\bibitem[{{Einaudi} {et~al.}(1996){Einaudi}, {Velli}, {Politano}, \&
  {Pouquet}}]{Einaudi1996}
{Einaudi}, G., {Velli}, M., {Politano}, H., \& {Pouquet}, A. 1996, \apjl, 457,
  L113, \dodoi{10.1086/309893}

\bibitem[{{Feinstein} {et~al.}(2020{\natexlab{a}}){Feinstein}, {Montet}, \&
  {Ansdell}}]{feinstein20_joss}
{Feinstein}, A., {Montet}, B., \& {Ansdell}, M. 2020{\natexlab{a}}, The Journal
  of Open Source Software, 5, 2347, \dodoi{10.21105/joss.02347}

\bibitem[{{Feinstein} {et~al.}(2020{\natexlab{b}}){Feinstein}, {Montet},
  {Ansdell}, {Nord}, {Bean}, {G{\"u}nther}, {Gully-Santiago}, \&
  {Schlieder}}]{feinstein20}
{Feinstein}, A.~D., {Montet}, B.~T., {Ansdell}, M., {et~al.}
  2020{\natexlab{b}}, \aj, 160, 219, \dodoi{10.3847/1538-3881/abac0a}

\bibitem[{{Feinstein} {et~al.}(2021){Feinstein}, {Seligman}, {G{\"u}nther}, \&
  {Adams}}]{Feinstein2021}
{Feinstein}, A.~D., {Seligman}, D.~Z., {G{\"u}nther}, M.~N., \& {Adams}, F.~C.
  2021, arXiv e-prints, arXiv:2109.07011.
\newblock \doarXiv{2109.07011}

\bibitem[{{Foreman-Mackey} {et~al.}(2013){Foreman-Mackey}, {Hogg}, {Lang}, \&
  {Goodman}}]{Foreman-Mackey13a}
{Foreman-Mackey}, D., {Hogg}, D.~W., {Lang}, D., \& {Goodman}, J. 2013, \pasp,
  125, 306, \dodoi{10.1086/670067}

\bibitem[{{Galsgaard} \& {Nordlund}(1996)}]{Galsgaard1996}
{Galsgaard}, K., \& {Nordlund}, {\r{A}}. 1996, \jgr, 101, 13445,
  \dodoi{10.1029/96JA00428}

\bibitem[{{Galtier}(1999)}]{Galtier1999}
{Galtier}, S. 1999, \apj, 521, 483, \dodoi{10.1086/307537}

\bibitem[{{Galtier} \& {Pouquet}(1998)}]{Galtier1998}
{Galtier}, S., \& {Pouquet}, A. 1998, \solphys, 179, 141,
  \dodoi{10.1023/A:1005056102064}

\bibitem[{{Georgoulis} {et~al.}(1998){Georgoulis}, {Velli}, \&
  {Einaudi}}]{Georgoulis1998}
{Georgoulis}, M.~K., {Velli}, M., \& {Einaudi}, G. 1998, \apj, 497, 957,
  \dodoi{10.1086/305486}

\bibitem[{{Ginsburg} {et~al.}(2019){Ginsburg}, {Sip{\H{o}}cz}, {Brasseur},
  {Cowperthwaite}, {Craig}, {Deil}, {Guillochon}, {Guzman}, {Liedtke}, {Lian
  Lim}, {Lockhart}, {Mommert}, {Morris}, {Norman}, {Parikh}, {Persson},
  {Robitaille}, {Segovia}, {Singer}, {Tollerud}, {de Val-Borro}, {Valtchanov},
  {Woillez}, {Astroquery Collaboration}, \& {a subset of astropy
  Collaboration}}]{astroquery19}
{Ginsburg}, A., {Sip{\H{o}}cz}, B.~M., {Brasseur}, C.~E., {et~al.} 2019, \aj,
  157, 98, \dodoi{10.3847/1538-3881/aafc33}

\bibitem[{{Golub} \& {Pasachoff}(1997)}]{Golub1997}
{Golub}, L., \& {Pasachoff}, J.~M. 1997, {The Solar Corona}

\bibitem[{{Goodman} \& {Weare}(2010)}]{goodman10}
{Goodman}, J., \& {Weare}, J. 2010, Communications in Applied Mathematics and
  Computational Science, 5, 65, \dodoi{10.2140/camcos.2010.5.65}

\bibitem[{{Greer} {et~al.}(2016){Greer}, {Hindman}, \& {Toomre}}]{Greer2016}
{Greer}, B.~J., {Hindman}, B.~W., \& {Toomre}, J. 2016, \apj, 824, 128,
  \dodoi{10.3847/0004-637X/824/2/128}

\bibitem[{{Grieger}(1992)}]{Grieger1992}
{Grieger}, B. 1992, Physica A Statistical Mechanics and its Applications, 191,
  51, \dodoi{10.1016/0378-4371(92)90505-K}

\bibitem[{{G{\"u}nther} {et~al.}(2020){G{\"u}nther}, {Zhan}, {Seager},
  {Rimmer}, {Ranjan}, {Stassun}, {Oelkers}, {Daylan}, {Newton}, {Kristiansen},
  {Olah}, {Gillen}, {Rappaport}, {Ricker}, {Vanderspek}, {Latham}, {Winn},
  {Jenkins}, {Glidden}, {Fausnaugh}, {Levine}, {Dittmann}, {Quinn},
  {Krishnamurthy}, \& {Ting}}]{guenther19_flares}
{G{\"u}nther}, M.~N., {Zhan}, Z., {Seager}, S., {et~al.} 2020, The Astronomical
  Journal, 159, 60, \dodoi{10.3847/1538-3881/ab5d3a}

\bibitem[{{Gutenberg} \& {Richter}(1956)}]{Gutenberg1956}
{Gutenberg}, B., \& {Richter}, C.~F. 1956, Annals of Geophysics, 53,
  \dodoi{10.4401/ag-4588}

\bibitem[{{Hale} {et~al.}(1919){Hale}, {Ellerman}, {Nicholson}, \&
  {Joy}}]{Hale1919}
{Hale}, G.~E., {Ellerman}, F., {Nicholson}, S.~B., \& {Joy}, A.~H. 1919, \apj,
  49, 153, \dodoi{10.1086/142452}

\bibitem[{Harris {et~al.}(2020)Harris, Millman, van~der Walt, Gommers,
  Virtanen, Cournapeau, Wieser, Taylor, Berg, Smith, Kern, Picus, Hoyer, van
  Kerkwijk, Brett, Haldane, del R{\'\i}o, Wiebe, Peterson, G{\'e}rard-Marchant,
  Sheppard, Reddy, Weckesser, Abbasi, Gohlke, \& Oliphant}]{numpy2020}
Harris, C.~R., Millman, K.~J., van~der Walt, S.~J., {et~al.} 2020, Nature, 585,
  357, \dodoi{10.1038/s41586-020-2649-2}

\bibitem[{{Hartmann} \& {Noyes}(1987)}]{hartmann87}
{Hartmann}, L.~W., \& {Noyes}, R.~W. 1987, \araa, 25, 271,
  \dodoi{10.1146/annurev.aa.25.090187.001415}

\bibitem[{Hesse \& Gross(2014)}]{Hesse2014}
Hesse, J., \& Gross, T. 2014, Frontiers in Systems Neuroscience, 8, 166,
  \dodoi{10.3389/fnsys.2014.00166}

\bibitem[{{Holzwarth}(2007)}]{Holzwarth2007}
{Holzwarth}, V. 2007, \memsai, 78, 271.
\newblock \doarXiv{0709.3008}

\bibitem[{{Howard}(1993)}]{Howard1993}
{Howard}, R.~F. 1993, \solphys, 145, 105, \dodoi{10.1007/BF00627986}

\bibitem[{{Howard} {et~al.}(2019){Howard}, {Corbett}, {Law}, {Ratzloff},
  {Glazier}, {Fors}, {del Ser}, \& {Haislip}}]{howard_ward19}
{Howard}, W.~S., {Corbett}, H., {Law}, N.~M., {et~al.} 2019, The Astrophysical
  Journal, 881, 9, \dodoi{10.3847/1538-4357/ab2767}

\bibitem[{{Howard} {et~al.}(2020){Howard}, {Corbett}, {Law}, {Ratzloff},
  {Galliher}, {Glazier}, {Fors}, {del Ser}, \& {Haislip}}]{howard20}
---. 2020, \apj, 895, 140, \dodoi{10.3847/1538-4357/ab9081}

\bibitem[{{Howell} {et~al.}(2014){Howell}, {Sobeck}, {Haas}, {Still},
  {Barclay}, {Mullally}, {Troeltzsch}, {Aigrain}, {Bryson}, {Caldwell},
  {Chaplin}, {Cochran}, {Huber}, {Marcy}, {Miglio}, {Najita}, {Smith},
  {Twicken}, \& {Fortney}}]{Howell2014}
{Howell}, S.~B., {Sobeck}, C., {Haas}, M., {et~al.} 2014, \pasp, 126, 398,
  \dodoi{10.1086/676406}

\bibitem[{Hunter(2007)}]{matplotlib}
Hunter, J.~D. 2007, Computing in Science \& Engineering, 9, 90,
  \dodoi{10.1109/MCSE.2007.55}

\bibitem[{{Ilin} {et~al.}(2019){Ilin}, {Schmidt}, {Davenport}, \&
  {Strassmeier}}]{ilin19}
{Ilin}, E., {Schmidt}, S.~J., {Davenport}, J. R.~A., \& {Strassmeier}, K.~G.
  2019, \aap, 622, A133, \dodoi{10.1051/0004-6361/201834400}

\bibitem[{{Ilin} {et~al.}(2021){Ilin}, {Schmidt}, {Poppenh{\"a}ger},
  {Davenport}, {Kristiansen}, \& {Omohundro}}]{ilin21}
{Ilin}, E., {Schmidt}, S.~J., {Poppenh{\"a}ger}, K., {et~al.} 2021, \aap, 645,
  A42, \dodoi{10.1051/0004-6361/202039198}

\bibitem[{{Ionson}(1982)}]{Ionson1982}
{Ionson}, J.~A. 1982, \apj, 254, 318, \dodoi{10.1086/159736}

\bibitem[{{Ionson}(1985)}]{Ionson1985}
---. 1985, \solphys, 100, 289, \dodoi{10.1007/BF00158433}

\bibitem[{{Kadanoff} {et~al.}(1989){Kadanoff}, {Nagel}, {Wu}, \&
  {Zhou}}]{Kadanoff1989}
{Kadanoff}, L.~P., {Nagel}, S.~R., {Wu}, L., \& {Zhou}, S.-M. 1989, \pra, 39,
  6524, \dodoi{10.1103/PhysRevA.39.6524}

\bibitem[{{K{\H{o}}v{\'a}ri} {et~al.}(2020){K{\H{o}}v{\'a}ri}, {Ol{\'a}h},
  {G{\"u}nther}, {Vida}, {Kriskovics}, {Seli}, {Bakos}, {Hartman}, {Csubry}, \&
  {Bhatti}}]{Khovaris2020}
{K{\H{o}}v{\'a}ri}, Z., {Ol{\'a}h}, K., {G{\"u}nther}, M.~N., {et~al.} 2020,
  \aap, 641, A83, \dodoi{10.1051/0004-6361/202038397}

\bibitem[{{Kitchatinov} \& {Olemskoy}(2015)}]{Kitchatinov2015}
{Kitchatinov}, L.~L., \& {Olemskoy}, S.~V. 2015, Research in Astronomy and
  Astrophysics, 15, 1801, \dodoi{10.1088/1674-4527/15/11/003}

\bibitem[{{Krucker} \& {Benz}(2000)}]{Krucker2000}
{Krucker}, S., \& {Benz}, A.~O. 2000, \solphys, 191, 341,
  \dodoi{10.1023/A:1005255608792}

\bibitem[{{Kulsrud}(1998)}]{Kulsrud1998}
{Kulsrud}, R.~M. 1998, Physics of Plasmas, 5, 1599, \dodoi{10.1063/1.872827}

\bibitem[{{Lee} {et~al.}(1993){Lee}, {Petrosian}, \& {McTiernan}}]{Lee1993}
{Lee}, T.~T., {Petrosian}, V., \& {McTiernan}, J.~M. 1993, \apj, 412, 401,
  \dodoi{10.1086/172929}

\bibitem[{{Lightkurve Collaboration} {et~al.}(2018){Lightkurve Collaboration},
  {Cardoso}, {Hedges}, {Gully-Santiago}, {Saunders}, {Cody}, {Barclay}, {Hall},
  {Sagear}, {Turtelboom}, {Zhang}, {Tzanidakis}, {Mighell}, {Coughlin}, {Bell},
  {Berta-Thompson}, {Williams}, {Dotson}, \& {Barentsen}}]{lightkurve}
{Lightkurve Collaboration}, {Cardoso}, J.~V.~d.~M., {Hedges}, C., {et~al.}
  2018, {Lightkurve: Kepler and TESS time series analysis in Python},
  Astrophysics Source Code Library.
\newblock \doeprint{1812.013}

\bibitem[{{Lin} {et~al.}(2019){Lin}, {Ip}, {Hou}, {Huang}, \&
  {Chang}}]{lin2019}
{Lin}, C.~L., {Ip}, W.~H., {Hou}, W.~C., {Huang}, L.~C., \& {Chang}, H.~Y.
  2019, The Astrophysical Journal, 873, 97, \dodoi{10.3847/1538-4357/ab041c}

\bibitem[{{Lin} {et~al.}(1984){Lin}, {Schwartz}, {Kane}, {Pelling}, \&
  {Hurley}}]{Lin1984}
{Lin}, R.~P., {Schwartz}, R.~A., {Kane}, S.~R., {Pelling}, R.~M., \& {Hurley},
  K.~C. 1984, \apj, 283, 421, \dodoi{10.1086/162321}

\bibitem[{{Litvinenko}(1996)}]{Litvinenko1996}
{Litvinenko}, Y.~E. 1996, \solphys, 167, 321, \dodoi{10.1007/BF00146342}

\bibitem[{Liu {et~al.}(2014)Liu, Hoeksema, Bobra, Hayashi, Schuck, \&
  Sun}]{Liu_2014_helicity_preference}
Liu, Y., Hoeksema, J.~T., Bobra, M., {et~al.} 2014, The Astrophysical Journal,
  785, 13, \dodoi{10.1088/0004-637x/785/1/13}

\bibitem[{{Longcope} \& {Fisher}(1996)}]{Longcope1996}
{Longcope}, D.~W., \& {Fisher}, G.~H. 1996, \apj, 458, 380,
  \dodoi{10.1086/176821}

\bibitem[{{Longcope} {et~al.}(1998){Longcope}, {Fisher}, \&
  {Pevtsov}}]{Longcope1998}
{Longcope}, D.~W., {Fisher}, G.~H., \& {Pevtsov}, A.~A. 1998, \apj, 507, 417,
  \dodoi{10.1086/306312}

\bibitem[{{Longcope} \& {Pevtsov}(2003)}]{Longcope2003}
{Longcope}, D.~W., \& {Pevtsov}, A.~A. 2003, Advances in Space Research, 32,
  1845, \dodoi{10.1016/S0273-1177(03)90618-1}

\bibitem[{{Longcope} {et~al.}(2007){Longcope}, {Ravindra}, \&
  {Barnes}}]{Longcope2007}
{Longcope}, D.~W., {Ravindra}, B., \& {Barnes}, G. 2007, \apj, 668, 571,
  \dodoi{10.1086/521095}

\bibitem[{{Longcope} \& {Sudan}(1994)}]{Longcope1994}
{Longcope}, D.~W., \& {Sudan}, R.~N. 1994, \apj, 437, 491,
  \dodoi{10.1086/175013}

\bibitem[{{Longcope} \& {Welsch}(2000)}]{Longcope2000}
{Longcope}, D.~W., \& {Welsch}, B.~T. 2000, \apj, 545, 1089,
  \dodoi{10.1086/317846}

\bibitem[{{Lu} \& {Hamilton}(1991)}]{lu1991}
{Lu}, E.~T., \& {Hamilton}, R.~J. 1991, \apjl, 380, L89, \dodoi{10.1086/186180}

\bibitem[{{Lu} {et~al.}(1993){Lu}, {Hamilton}, {McTiernan}, \&
  {Bromund}}]{Lu1993}
{Lu}, E.~T., {Hamilton}, R.~J., {McTiernan}, J.~M., \& {Bromund}, K.~R. 1993,
  \apj, 412, 841, \dodoi{10.1086/172966}

\bibitem[{{Lu} {et~al.}(2020){Lu}, {Angus}, {Ag{\"u}eros}, {Blancato}, {Ness},
  {Rowland}, {Curtis}, \& {Grunblatt}}]{lu2020}
{Lu}, Y., {Angus}, R., {Ag{\"u}eros}, M.~A., {et~al.} 2020, \aj, 160, 168,
  \dodoi{10.3847/1538-3881/abada4}

\bibitem[{{MacTaggart} {et~al.}(2021){MacTaggart}, {Prior}, {Raphaldini},
  {Romano}, \& {Guglielmino}}]{MacTaggart2021}
{MacTaggart}, D., {Prior}, C., {Raphaldini}, B., {Romano}, P., \&
  {Guglielmino}, S. 2021, arXiv e-prints, arXiv:2106.11638.
\newblock \doarXiv{2106.11638}

\bibitem[{{Maggio} {et~al.}(1987){Maggio}, {Sciortino}, {Vaiana}, {Majer},
  {Bookbinder}, {Golub}, {Harnden}, \& {Rosner}}]{maggio87}
{Maggio}, A., {Sciortino}, S., {Vaiana}, G.~S., {et~al.} 1987, \apj, 315, 687,
  \dodoi{10.1086/165170}

\bibitem[{{Mamajek} \& {Hillenbrand}(2008)}]{Mamajek2008}
{Mamajek}, E.~E., \& {Hillenbrand}, L.~A. 2008, \apj, 687, 1264,
  \dodoi{10.1086/591785}

\bibitem[{{Manek} \& {Brummell}(2021)}]{Manek2021}
{Manek}, B., \& {Brummell}, N. 2021, \apj, 909, 72,
  \dodoi{10.3847/1538-4357/abd859}

\bibitem[{{Manek} {et~al.}(2018){Manek}, {Brummell}, \& {Lee}}]{Manek2018}
{Manek}, B., {Brummell}, N., \& {Lee}, D. 2018, \apjl, 859, L27,
  \dodoi{10.3847/2041-8213/aac723}

\bibitem[{{Mikic} {et~al.}(1989){Mikic}, {Schnack}, \& {van Hoven}}]{mikic1989}
{Mikic}, Z., {Schnack}, D.~D., \& {van Hoven}, G. 1989, \apj, 338, 1148,
  \dodoi{10.1086/167265}

\bibitem[{{Mitra-Kraev} \& {Benz}(2001)}]{MitraKraev2001}
{Mitra-Kraev}, U., \& {Benz}, A.~O. 2001, \aap, 373, 318,
  \dodoi{10.1051/0004-6361:20010524}

\bibitem[{{Moffatt}(1978)}]{Moffat1978}
{Moffatt}, H.~K. 1978, {Magnetic field generation in electrically conducting
  fluids}

\bibitem[{{Moffatt} \& {Tsinober}(1990)}]{Moffatt1990}
{Moffatt}, H.~K., \& {Tsinober}, A. 1990, {Topological Fluid Mechanics}

\bibitem[{{Morales} \& {Santos}(2020)}]{Morales2020}
{Morales}, L.~F., \& {Santos}, N.~A. 2020, \solphys, 295, 155,
  \dodoi{10.1007/s11207-020-01713-0}

\bibitem[{{Nagel} \& {Herrmann}(1993)}]{Nagel1993}
{Nagel}, K., \& {Herrmann}, H.~J. 1993, Physica A Statistical Mechanics and its
  Applications, 199, 254, \dodoi{10.1016/0378-4371(93)90006-P}

\bibitem[{{Newman}(1996)}]{Newman1996_criticality}
{Newman}, M.~E.~J. 1996, Proceedings of the Royal Society of London Series B,
  263, 1605.
\newblock \doarXiv{adap-org/9607002}

\bibitem[{{Newman} \& {Sneppen}(1996)}]{Newman1996}
{Newman}, M.~E.~J., \& {Sneppen}, K. 1996, \pre, 54, 6226,
  \dodoi{10.1103/PhysRevE.54.6226}

\bibitem[{{Notsu} {et~al.}(2013){Notsu}, {Shibayama}, {Maehara}, {Notsu},
  {Nagao}, {Honda}, {Ishii}, {Nogami}, \& {Shibata}}]{Notsu2013}
{Notsu}, Y., {Shibayama}, T., {Maehara}, H., {et~al.} 2013, \apj, 771, 127,
  \dodoi{10.1088/0004-637X/771/2/127}

\bibitem[{{Noyes} {et~al.}(1984){Noyes}, {Hartmann}, {Baliunas}, {Duncan}, \&
  {Vaughan}}]{Noyes1984}
{Noyes}, R.~W., {Hartmann}, L.~W., {Baliunas}, S.~L., {Duncan}, D.~K., \&
  {Vaughan}, A.~H. 1984, \apj, 279, 763, \dodoi{10.1086/161945}

\bibitem[{Olami {et~al.}(1992)Olami, Feder, \& Christensen}]{Olami1992}
Olami, Z., Feder, H. J.~S., \& Christensen, K. 1992, Phys. Rev. Lett., 68,
  1244, \dodoi{10.1103/PhysRevLett.68.1244}

\bibitem[{{Osten} \& {Brown}(1999)}]{Osten1999}
{Osten}, R.~A., \& {Brown}, A. 1999, \apj, 515, 746, \dodoi{10.1086/307034}

\bibitem[{{Pallavicini} {et~al.}(1981){Pallavicini}, {Golub}, {Rosner},
  {Vaiana}, {Ayres}, \& {Linsky}}]{Pallavicini1981}
{Pallavicini}, R., {Golub}, L., {Rosner}, R., {et~al.} 1981, \apj, 248, 279,
  \dodoi{10.1086/159152}

\bibitem[{{Parker}(1955)}]{Parker1955}
{Parker}, E.~N. 1955, \apj, 122, 293, \dodoi{10.1086/146087}

\bibitem[{{Parker}(1972)}]{Parker1972}
---. 1972, \apj, 174, 499, \dodoi{10.1086/151512}

\bibitem[{{Parker}(1983)}]{Parker1983}
---. 1983, \apj, 264, 642, \dodoi{10.1086/160637}

\bibitem[{{Parker}(1988)}]{Parker1988}
---. 1988, \apj, 330, 474, \dodoi{10.1086/166485}

\bibitem[{{Parker}(1989)}]{Parker1989}
---. 1989, \solphys, 121, 271, \dodoi{10.1007/BF00161700}

\bibitem[{{Parker}(1994)}]{Parker1994}
---. 1994, Spontaneous current sheets in magnetic fields : with applications to
  stellar x-rays. International Series in Astronomy and Astrophysics, 1

\bibitem[{{Parker}(2009)}]{Parker2009}
---. 2009, \ssr, 144, 15, \dodoi{10.1007/s11214-008-9445-x}

\bibitem[{{Pevtsov} {et~al.}(1995){Pevtsov}, {Canfield}, \&
  {Metcalf}}]{Pevtsov1995}
{Pevtsov}, A.~A., {Canfield}, R.~C., \& {Metcalf}, T.~R. 1995, \apjl, 440,
  L109, \dodoi{10.1086/187773}

\bibitem[{{Porter} {et~al.}(1987){Porter}, {Moore}, {Reichmann}, {Engvold}, \&
  {Harvey}}]{Porter1987}
{Porter}, J.~G., {Moore}, R.~L., {Reichmann}, E.~J., {Engvold}, O., \&
  {Harvey}, K.~L. 1987, \apj, 323, 380, \dodoi{10.1086/165835}

\bibitem[{{Price-Whelan} {et~al.}(2018){Price-Whelan}, {Sip{\H{o}}cz},
  {G{\"u}nther}, {Lim}, {Crawford}, {Conseil}, {Shupe}, {Craig}, {Dencheva},
  {Ginsburg}, {VanderPlas}, {Bradley}, {P{\'e}rez-Su{\'a}rez}, {de Val-Borro},
  {Paper Contributors}, {Aldcroft}, {Cruz}, {Robitaille}, {Tollerud},
  {Coordination Committee}, {Ardelean}, {Babej}, {Bach}, {Bachetti}, {Bakanov},
  {Bamford}, {Barentsen}, {Barmby}, {Baumbach}, {Berry}, {Biscani}, {Boquien},
  {Bostroem}, {Bouma}, {Brammer}, {Bray}, {Breytenbach}, {Buddelmeijer},
  {Burke}, {Calderone}, {Cano Rodr{\'\i}guez}, {Cara}, {Cardoso}, {Cheedella},
  {Copin}, {Corrales}, {Crichton}, {D{\textquoteright}Avella}, {Deil},
  {Depagne}, {Dietrich}, {Donath}, {Droettboom}, {Earl}, {Erben}, {Fabbro},
  {Ferreira}, {Finethy}, {Fox}, {Garrison}, {Gibbons}, {Goldstein}, {Gommers},
  {Greco}, {Greenfield}, {Groener}, {Grollier}, {Hagen}, {Hirst}, {Homeier},
  {Horton}, {Hosseinzadeh}, {Hu}, {Hunkeler}, {Ivezi{\'c}}, {Jain}, {Jenness},
  {Kanarek}, {Kendrew}, {Kern}, {Kerzendorf}, {Khvalko}, {King}, {Kirkby},
  {Kulkarni}, {Kumar}, {Lee}, {Lenz}, {Littlefair}, {Ma}, {Macleod},
  {Mastropietro}, {McCully}, {Montagnac}, {Morris}, {Mueller}, {Mumford},
  {Muna}, {Murphy}, {Nelson}, {Nguyen}, {Ninan}, {N{\"o}the}, {Ogaz}, {Oh},
  {Parejko}, {Parley}, {Pascual}, {Patil}, {Patil}, {Plunkett}, {Prochaska},
  {Rastogi}, {Reddy Janga}, {Sabater}, {Sakurikar}, {Seifert}, {Sherbert},
  {Sherwood-Taylor}, {Shih}, {Sick}, {Silbiger}, {Singanamalla}, {Singer},
  {Sladen}, {Sooley}, {Sornarajah}, {Streicher}, {Teuben}, {Thomas},
  {Tremblay}, {Turner}, {Terr{\'o}n}, {van Kerkwijk}, {de la Vega}, {Watkins},
  {Weaver}, {Whitmore}, {Woillez}, {Zabalza}, \& {Contributors}}]{astropy18}
{Price-Whelan}, A.~M., {Sip{\H{o}}cz}, B.~M., {G{\"u}nther}, H.~M., {et~al.}
  2018, \aj, 156, 123, \dodoi{10.3847/1538-3881/aabc4f}

\bibitem[{{Priest} \& {Forbes}(2000)}]{Priest2000}
{Priest}, E., \& {Forbes}, T. 2000, {Magnetic Reconnection}

\bibitem[{{Prior} \& {MacTaggart}(2016)}]{Prior2016}
{Prior}, C., \& {MacTaggart}, D. 2016, Geophysical and Astrophysical Fluid
  Dynamics, 110, 432, \dodoi{10.1080/03091929.2016.1216552}

\bibitem[{{Raetz} {et~al.}(2020){Raetz}, {Stelzer}, {Damasso}, \&
  {Scholz}}]{raetz20}
{Raetz}, S., {Stelzer}, B., {Damasso}, M., \& {Scholz}, A. 2020, Astronomy and
  Astrophysics, 637, A22, \dodoi{10.1051/0004-6361/201937350}

\bibitem[{{Rempel}(2011)}]{Rempel2011}
{Rempel}, M. 2011, {Solar Convection Zone Dynamics}, ed. M.~P. {Miralles} \&
  J.~{S{\'a}nchez Almeida}, Vol.~4, 23

\bibitem[{{Ribeiro} {et~al.}(2010){Ribeiro}, {Copelli}, {Caixeta}, {Belchior},
  {Chialvo}, {Nicolelis}, \& {Ribeiro}}]{Ribeiro2010}
{Ribeiro}, T.~L., {Copelli}, M., {Caixeta}, F., {et~al.} 2010, PLoS ONE, 5,
  e14129, \dodoi{10.1371/journal.pone.0014129}

\bibitem[{{Ricker} {et~al.}(2015){Ricker}, {Winn}, {Vanderspek}, {Latham},
  {Bakos}, {Bean}, {Berta-Thompson}, {Brown}, {Buchhave}, {Butler}, {Butler},
  {Chaplin}, {Charbonneau}, {Christensen-Dalsgaard}, {Clampin}, {Deming},
  {Doty}, {De Lee}, {Dressing}, {Dunham}, {Endl}, {Fressin}, {Ge}, {Henning},
  {Holman}, {Howard}, {Ida}, {Jenkins}, {Jernigan}, {Johnson}, {Kaltenegger},
  {Kawai}, {Kjeldsen}, {Laughlin}, {Levine}, {Lin}, {Lissauer}, {MacQueen},
  {Marcy}, {McCullough}, {Morton}, {Narita}, {Paegert}, {Palle}, {Pepe},
  {Pepper}, {Quirrenbach}, {Rinehart}, {Sasselov}, {Sato}, {Seager},
  {Sozzetti}, {Stassun}, {Sullivan}, {Szentgyorgyi}, {Torres}, {Udry}, \&
  {Villasenor}}]{ricker15}
{Ricker}, G.~R., {Winn}, J.~N., {Vanderspek}, R., {et~al.} 2015, Journal of
  Astronomical Telescopes, Instruments, and Systems, 1, 014003,
  \dodoi{10.1117/1.JATIS.1.1.014003}

\bibitem[{{Rosner} \& {Vaiana}(1978)}]{Rosner1978}
{Rosner}, R., \& {Vaiana}, G.~S. 1978, \apj, 222, 1104, \dodoi{10.1086/156227}

\bibitem[{{Schrijver} \& {Zwaan}(2000)}]{Schrijver2000}
{Schrijver}, C.~J., \& {Zwaan}, C. 2000, {Solar and Stellar Magnetic Activity}

\bibitem[{{Seehafer}(1990)}]{Seehafer1990}
{Seehafer}, N. 1990, \solphys, 125, 219, \dodoi{10.1007/BF00158402}

\bibitem[{{Seligman} {et~al.}(2014){Seligman}, {Petrie}, \&
  {Komm}}]{Seligman2014}
{Seligman}, D., {Petrie}, G.~J.~D., \& {Komm}, R. 2014, \apj, 795, 113,
  \dodoi{10.1088/0004-637X/795/2/113}

\bibitem[{{Shakhovskaia}(1989)}]{Shakhovskaia1989}
{Shakhovskaia}, N.~I. 1989, \solphys, 121, 375, \dodoi{10.1007/BF00161707}

\bibitem[{{Shibayama} {et~al.}(2013){Shibayama}, {Maehara}, {Notsu}, {Notsu},
  {Nagao}, {Honda}, {Ishii}, {Nogami}, \& {Shibata}}]{shibayama13}
{Shibayama}, T., {Maehara}, H., {Notsu}, S., {et~al.} 2013, The Astrophysical
  Journals, 209, 5, \dodoi{10.1088/0067-0049/209/1/5}

\bibitem[{{Soderblom}(2010)}]{Soderblom2010}
{Soderblom}, D.~R. 2010, \araa, 48, 581,
  \dodoi{10.1146/annurev-astro-081309-130806}

\bibitem[{{Solanki} {et~al.}(2006){Solanki}, {Inhester}, \&
  {Sch{\"u}ssler}}]{Solanki2006}
{Solanki}, S.~K., {Inhester}, B., \& {Sch{\"u}ssler}, M. 2006, Reports on
  Progress in Physics, 69, 563, \dodoi{10.1088/0034-4885/69/3/R02}

\bibitem[{{Sornette} \& {Sornette}(1989)}]{Sornette1989}
{Sornette}, A., \& {Sornette}, D. 1989, EPL (Europhysics Letters), 9, 197,
  \dodoi{10.1209/0295-5075/9/3/002}

\bibitem[{{Sturrock} {et~al.}(1990){Sturrock}, {Dixon}, {Klimchuk}, \&
  {Antiochos}}]{Sturrock1990}
{Sturrock}, P.~A., {Dixon}, W.~W., {Klimchuk}, J.~A., \& {Antiochos}, S.~K.
  1990, \apjl, 356, L31, \dodoi{10.1086/185743}

\bibitem[{{Sturrock} {et~al.}(1984){Sturrock}, {Kaufman}, {Moore}, \&
  {Smith}}]{Sturrock1984}
{Sturrock}, P.~A., {Kaufman}, P., {Moore}, R.~L., \& {Smith}, D.~F. 1984,
  \solphys, 94, 341, \dodoi{10.1007/BF00151322}

\bibitem[{{Temme}(1975)}]{Temme1975}
{Temme}, N.~M. 1975, Mathematics of Computation, 29, 1109,
  \dodoi{10.2307/2005750}

\bibitem[{{Turcotte}(1999)}]{Turcotte1999}
{Turcotte}, D.~L. 1999, Reports on Progress in Physics, 62, 1377,
  \dodoi{10.1088/0034-4885/62/10/201}

\bibitem[{Turcotte {et~al.}(2002)Turcotte, Malamud, Guzzetti, \&
  Reichenbach}]{Turcotte2002}
Turcotte, D.~L., Malamud, B.~D., Guzzetti, F., \& Reichenbach, P. 2002,
  Proceedings of the National Academy of Sciences, 99, 2530,
  \dodoi{10.1073/pnas.012582199}

\bibitem[{{van Ballegooijen}(1986)}]{vanBallegooijen1986}
{van Ballegooijen}, A.~A. 1986, \apj, 311, 1001, \dodoi{10.1086/164837}

\bibitem[{{van Ballegooijen} {et~al.}(2014){van Ballegooijen}, {Asgari-Targhi},
  \& {Berger}}]{vanBallegooijen2014}
{van Ballegooijen}, A.~A., {Asgari-Targhi}, M., \& {Berger}, M.~A. 2014, \apj,
  787, 87, \dodoi{10.1088/0004-637X/787/1/87}

\bibitem[{{van Ballegooijen} {et~al.}(2011){van Ballegooijen}, {Asgari-Targhi},
  {Cranmer}, \& {DeLuca}}]{vanBallegooijen2011}
{van Ballegooijen}, A.~A., {Asgari-Targhi}, M., {Cranmer}, S.~R., \& {DeLuca},
  E.~E. 2011, \apj, 736, 3, \dodoi{10.1088/0004-637X/736/1/3}

\bibitem[{Van Der~Walt {et~al.}(2011)Van Der~Walt, Colbert, \&
  Varoquaux}]{numpy}
Van Der~Walt, S., Colbert, S.~C., \& Varoquaux, G. 2011, Computing in Science
  \& Engineering, 13, 22

\bibitem[{{Vilhu}(1984)}]{Vilhu1984}
{Vilhu}, O. 1984, \aap, 133, 117

\bibitem[{{Weber} \& {Browning}(2016)}]{Weber2016}
{Weber}, M.~A., \& {Browning}, M.~K. 2016, \apj, 827, 95,
  \dodoi{10.3847/0004-637X/827/2/95}

\bibitem[{{Weber} {et~al.}(2011){Weber}, {Fan}, \& {Miesch}}]{Weber2011}
{Weber}, M.~A., {Fan}, Y., \& {Miesch}, M.~S. 2011, \apj, 741, 11,
  \dodoi{10.1088/0004-637X/741/1/11}

\bibitem[{{Weber} {et~al.}(2013){Weber}, {Fan}, \& {Miesch}}]{Weber2013}
---. 2013, \solphys, 287, 239, \dodoi{10.1007/s11207-012-0093-7}

\bibitem[{{Wheatland}(2000)}]{Wheatland2000}
{Wheatland}, M.~S. 2000, \apj, 532, 1209, \dodoi{10.1086/308605}

\bibitem[{{Withbroe} \& {Noyes}(1977)}]{Withbroe1977}
{Withbroe}, G.~L., \& {Noyes}, R.~W. 1977, \araa, 15, 363,
  \dodoi{10.1146/annurev.aa.15.090177.002051}

\bibitem[{{Wright} \& {Drake}(2016)}]{Wright2016}
{Wright}, N.~J., \& {Drake}, J.~J. 2016, \nat, 535, 526,
  \dodoi{10.1038/nature18638}

\bibitem[{{Wright} {et~al.}(2011){Wright}, {Drake}, {Mamajek}, \&
  {Henry}}]{wright11}
{Wright}, N.~J., {Drake}, J.~J., {Mamajek}, E.~E., \& {Henry}, G.~W. 2011,
  \apj, 743, 48, \dodoi{10.1088/0004-637X/743/1/48}

\bibitem[{{Yang} \& {Liu}(2019)}]{yang19}
{Yang}, H., \& {Liu}, J. 2019, The Astrophysical Journals, 241, 29,
  \dodoi{10.3847/1538-4365/ab0d28}

\bibitem[{{Zirker} \& {Cleveland}(1993{\natexlab{a}})}]{Zirker1993i}
{Zirker}, J.~B., \& {Cleveland}, F.~M. 1993{\natexlab{a}}, \solphys, 144, 341,
  \dodoi{10.1007/BF00627598}

\bibitem[{{Zirker} \& {Cleveland}(1993{\natexlab{b}})}]{Zirker1993ii}
---. 1993{\natexlab{b}}, \solphys, 145, 119, \dodoi{10.1007/BF00627988}

\end{thebibliography}
\bibliographystyle{aasjournal}

\end{document}